\documentclass[pra,superscriptaddress,twocolumn,showpacs]{revtex4-2} 
\usepackage{graphicx}
\usepackage{amsmath,amssymb,bm,mathtools} 
\usepackage{textcomp} 
\usepackage{xspace} 
\usepackage{soul,hyperref}
\usepackage[usenames,dvipsnames]{color}
\usepackage[caption=false]{subfig}
\usepackage{pstool}
\usepackage[caption=false]{subfig}
\usepackage{siunitx,physics}
\DeclareSIUnit\gauss{G}

\newcommand{\iso}[1]{\ensuremath{{^{#1}}}}
\newcommand{\tplus}{\ensuremath{^{3+}}\xspace}

\newcommand{\erg}{\ensuremath{^4}I\ensuremath{_{15/2}}\xspace}
\newcommand{\ere}{\ensuremath{^4}I\ensuremath{_{13/2}}\xspace}

\newcommand{\note}[1]{}

\newcommand{\citeit}[1]{\cite{#1}\xspace} 

\begin{document}

\title{Optical and Zeeman spectroscopy of individual Er ion pairs in silicon}
\author{Guangchong \surname{Hu}}

\affiliation{Centre  for Quantum Computation and Communication Technology, School of Physics, University of New South Wales, Australia}
\author{Rose L. \surname{Ahlefeldt}}

\email{rose.ahlefeldt@anu.edu.au} 
\affiliation{Centre for Quantum Computation and Communication Technology, The Australian National University, Australia}

\author{Gabriele G. \surname{de Boo}}
\affiliation{Centre of Excellence for Quantum Computation and Communication Technology, School of Physics, University of New South Wales, Australia}

\author{Alexey \surname{Lyasota}}
\affiliation{Centre for Quantum Computation and Communication Technology, School of Physics, University of New South Wales, Australia}

\author{Brett C. \surname{Johnson}}
\affiliation{Centre for Quantum Computation and Communication Technology, RMIT University, Australia}
\affiliation{Centre for Quantum Computation and Communication Technology, The University of Melbourne, Australia}
\author{Jeffrey C. \surname{McCallum}}
\affiliation{Centre for Quantum Computation and Communication Technology, The University of Melbourne, Australia}

\author{Matthew J. \surname{Sellars}}
\affiliation{Centre for Quantum Computation and Communication Technology, The Australian National University, Australia}

\author{Chunming \surname{Yin}} 
\affiliation{Centre of Excellence for Quantum Computation and Communication Technology, School of Physics, University of New South Wales, Australia}

\affiliation{Hefei National Laboratory for Physical Sciences at the Microscale, CAS Key Laboratory of Microscale Magnetic Resonance and School of Physical Sciences, University of Science and Technology of China, China}

\author{Sven \surname{Rogge}}
\affiliation{Centre of Excellence for Quantum Computation and Communication Technology, School of Physics, University of New South Wales, Australia}

\date{\today}
\begin{abstract}
We make the first study the optical energy level structure and interactions of pairs of single rare earth ions using a hybrid electro-optical detection method applied to Er-implanted silicon. Two examples of Er\tplus pairs were identified in the optical spectrum by their characteristic energy level splitting patterns, and linear Zeeman spectra were used to characterise the sites. One pair is positively identified as two identical Er\tplus ions in sites of at least C$_2$ symmetry coupled via a large, 200~GHz Ising-like spin interaction and 1.5~GHz resonant optical interaction. Small non-Ising contributions to the spin interaction are attributed to distortion of the site measurable because of the high resolution of the single-ion measurement. The interactions are compared to previous measurements made using rare earth ensemble systems, and the application of this type of strongly coupled ion array to quantum computing is discussed.
\end{abstract}


\maketitle

\section{Introduction}
Optically coupled single spins in solid state hosts are studied for a range of quantum technologies, including quantum memories, repeaters, and computers  due to three main features: spin states with low decoherence, optical transitions to interface the solid state qubit with flying qubits, and a physical platform compatible with scalable device engineering \citeit{awschalom18}. For computing in particular, the possibility of creating arrays of closely spaced, optically coupled spins is attractive as these arrays can provide a multi-qubit system where the short-range interactions between different qubits enable fast multi-qubit gates. For instance, by using a set of \iso{13}C nuclear spins coupled to an optically active  diamond nitrogen vacancy center, entangling gate operations have been achieved in a ten-qubit array \citeit{bradley19}. 

While most work in this area has concentrated on bright optical color centers such as the nitrogen vacancy in diamond, rare earth ions in various solids are increasingly being studied. These materials have much weaker optical coupling strengths, but several groups have recently shown the detection \citeit{kolesov12,utikal14,nakamura14,kolesov13,yin13, groot-berning19} and coherent control \citeit{chen20,raha20,siyushev14, kindem20, kornher20} of single rare earth ions. Er\tplus is particularly interesting as it offers an optical transition between its \erg --\ere levels that lies within the \SI{1550}{\nano\metre} transition band of optical fibers, and crystals doped with Er\tplus have shown very long optical \citeit{bottger09} and spin \citeit{rancic18} coherence times in the absence of magnetic noise. Single Er\tplus ions have been observed using hybrid electrical-optical \citeit{yin13,deboo20} and optical-cavity enhanced detection \citeit{dibos18,raha20,chen20}. Six individual Er\tplus ions have been individually addressed in a single optical cavity \citeit{chen20}, but no arrays of Er ions have been studied to date.  

Here, we report the first measurements of individual interacting Er ion pairs. These measurements were made in silicon, which can be engineered into photonic devices using CMOS fabrication techniques, and is a low-spin host offering a quiet magnetic environment for the Er\tplus ions. However, the solubility of Er\tplus in silicon is low,  and Er is typically  included in the silicon lattice by ion implantation or during molecular beam epitaxy growth, often co-doped with oxygen or other light impurities to enhance the erbium-associated luminescence \citeit{carey98,priolo95, liu95}. Implanted Er\tplus in Si has shown comparable spin coherence times in low magnetic fields to Er substitutional dopants \citeit{hughes21}, but it can occupy a large variety of possible sites in silicon, and the properties of these sites, even for ensembles, are not well known. Sites have been studied with photoluminescence \citeit{przybylinska96, tang89, vinh03}, photoluminescence excitation \citeit{weiss21,berkman21}  and electron paramagnetic resonance \citeit{carey96, carey98,carey99,carey02}. However, only a  small number of sites have been positively identified,  typically as Er-O complexes of low symmetry, including multiple monoclinic \citeit{carey99} and trigonal \citeit{carey99} sites, and one orthorhombic site \citeit{vinh03}. Only a single bulk study \citeit{hughes19} has reported  a possible Er\tplus pair site, and no information on the interaction between the ions was obtained. 

\begin{figure}[htbp]
	\centering
	   	\includegraphics[width=1.0\columnwidth]{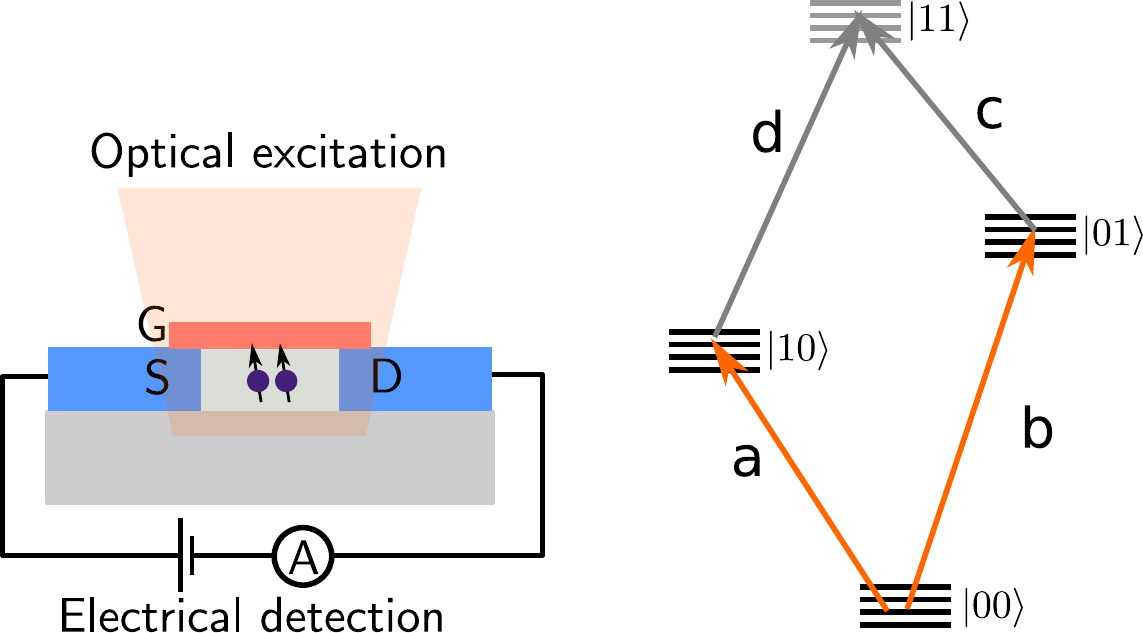}
	\caption[specB]{Concept of studying Er pair sites using hybrid electro-optic detection. Left: An Er ion pair sitting in the channel region underneath the gate (G) of a fin-field effect transistor (finFET) is optically excited from the (Kramers doublet) \erg ($\ket{0}$) state to the \ere ($\ket{1}$) state, and the non-radiative decay via nearby trap sites produces a current between source (S) and drain (D), forming the detected signal. Right: representation of the energy level structure of an Er\tplus ion pair, showing transitions probed. The coupled pair has four electronic states, $\ket{00}$, $\ket{10}$, $\ket{01}$, and $\ket{11}$, each composed of four electron spin levels for the $I=0$ isotopes studied. We excite only transitions "a" for the non-identical site and "a" and "b" for the identical site, shown in orange, although transitions "c" and "d" (gray) can in principle be excited via a two-laser process.  }
	\label{fig:concept}
\end{figure}
In this work, we study the \erg--\ere transition of Er ion pairs using a hybrid electrical-optical  detection method \citeit{yin13} (see Fig \ref{fig:concept}). We present spectra in zero and applied magnetic field for two different types of Er pairs sites: one composed of Er\tplus ions in non-identical sites, and a second of ions in identical sites. We analyze these spectra using a simplified model for spin and optical coupling between the two ions. This work is not only the first study of individual rare earth ion pairs in the solid state, but the first direct study of Er pair sites in silicon. The high resolution possible by studying a single instance of the pair, rather than an ensemble of pairs, allows precise determination of the properties of the individual Er ions as well as of the interaction between them, improving our understanding of rare earth interactions for future quantum computing applications \citeit{guillot-noel04}.

\section{Interactions between rare earth ion pairs}\label{sec:pairint}
While single rare earth ion pairs have not previously been studied, many workers have investigated pair interactions in ensembles where the rare earth ions are substitutional dopants. This work is reviewed by Cone and Meltzer \citeit{cone87} (focused on optical energy transfer) and, more recently, Guillot-No\"el et al. \citeit{guillot-noel04} (focused on spin interactions). The possible interactions between two ions in a pair can be separated into two  classes: indirect and direct interactions. The main indirect interaction is a strain shift arising from the crystal distortion due to the incorporation of dopant ions.  For an ion pair, this strain means the presence of the partner ion shifts, in particular, the optical transition frequency of each ion in the pair (either identically for ions in the same site, or differently for ions in different sites).  The shift is unique to the pair site, which has allowed site-selective measurements of pair interactions in ensemble systems. These strain shifts can vary from GHz to THz  (e.g. \citeit{ fricke79a, yamaguchi98, ahlefeldt13method, laplane16high}).

In a quantum computer, strain shifts would determine the optical frequency of a pair of qubits, but would play no role in gate operations. Thus, here we are concerned with direct interactions between the wavefunctions of the two ions of the pair. There are four main mechanisms considered for these interactions: exchange, magnetic multipole interactions, electric multipole interactions, and virtual phonon exchange \citeit{wolf71,cone87,baker71}. The relative contribution of these terms is strongly dependent on the interacting ions, and in principle accurately estimating it requires knowledge of the full electronic wavefunction of the individual ions. However, the interaction can often be split into an optical component, and a spin component, with the latter fitted directly to experimental data. This is the approach taken here. 

We will first briefly discuss the optical interaction. Modelling contributions to this part does require an accurate crystal field model, which was not available in our measurements as these sites were seen for the first time, and we studied only a two crystal field levels (ground and excited state). The off-diagonal part of the optical interaction leads to energy transfer, while the diagonal part causes a frequency shift on one ion when the other is excited. This latter ``blockade'' mechanism is utilized for multi-qubit gates in rare earth quantum computing proposals, where the mechanism can be electric dipole \citeit{ohlsson02}, magnetic dipole \citeit{grimm21} or mixed \citeit{ahlefeldt20}. In a single-laser experiment, this blockade effect is difficult to distinguish from the strain shift, but it can be measured in a two laser experiment, yielding values for nearest neighbors from, for instance, 50~MHz in EuCl$_3$.6H$_2$O \citeit{ahlefeldt13precision} to 5~GHz in Pr:LaF$_3$ \citeit{lukac89}.

\subsection{Spin interactions of ion pairs}
The electron or nuclear spin splitting of an isolated electronic state for a single rare earth ion can be accurately modelled using a reduced spin Hamiltonian. Such a model includes only these degrees of freedom, parametrizing any electronic contributions to the structure. The same model can be used to study spin interactions between ion pairs \cite{guillot-noel04}. Here, we describe the spin Hamiltonian model appropriate to the Er\tplus pairs we studied.

In low symmetry sites, the spin Hamiltonian of an individual Er ion is anisotropic and different in every electronic state. Therefore, the most general case we need to model for a pair of coupled ions is two optically active Er ions in different low symmetry sites. Then, we need to develop reduced Hamiltonians for three electronic levels: $\ket{00}$, with both ions in the ground state; $\ket{01}$, with ion 1 in the ground state, and ion 2 in the optical excited state; and $\ket{10}$, with ion 1 in the excited state, and ion 2 in the ground state. We ignore $\ket{11}$ because we have not seen signatures of this double-excitation in optical spectra. Assuming both Er ions are isotopes with nuclear spin $I=0$ (the case for the pairs studied here), the spin Hamiltonian for an electronic state can be written:
\begin{equation}
H_{mn} = \bm{B\cdot M^{1m}\cdot \hat{S}_1}+\bm{B\cdot M^{2n}\cdot \hat{S}_2}+\bm{\hat{S_1}\cdot J^{mn}\cdot \hat{S}_2}
\label{eq:hamil}
\end{equation}
for $mn = \ket{00},\ket{10},\ket{01}$. In this equation, the first two terms describe the Zeeman interactions of the two individual ions, and the third term gives the coupling between the two spins. Specifically, $\bm{B}$ is the external magnetic field,  $\bm{M^{mi}}$ is the anisotropic Zeeman tensor (g-tensor) of the ion $i$  ($i=1,2$) in state $m$, $\bm{\hat{S}_i}$ is the spin operator for ion $i$, and $\bm{J^{mn}}$ is the spin-spin interaction tensor for ion 1 in electronic state $\ket{m}$ and ion 2 in state $\ket{n}$. The spin operators $\bm{\hat{S}_i}$ are normal spin-half vector operators expressed in space of the coupled ions as:
\begin{align}
    \bm{\hat{S_1}} &\implies \bm{\hat{I}\otimes \hat{S_1}}\\\nonumber
    \bm{\hat{S_2}} &\implies \bm{\hat{S_2}\otimes \hat{I}}
\end{align}
for $\bm{\hat{I}}$ the identity operator.
In the general case the interaction between the two spins is completely anisotropic, with all nine terms of $\bm{J}$ independent. It is common to write this spin interaction as a sum of three terms, all of different shape:
\begin{align}\label{eq:shape}
    \bm{S_1\cdot J\cdot S_2} &= J_0\bm{S_1\cdot S_2}+\sum_{\alpha,\beta}V_{\alpha \beta}\left(S_1^\alpha S_2^\beta+S_1^\beta S_2^\alpha\right)\\ \nonumber &+\bm{D\cdot\left(S_1\times S_2\right)}
\end{align}
These terms are, in order, the isotropic term, the symmetric traceless term, and the antisymmetric term. This decomposition is particularly useful for certain spins, such as $3d$ ions with quenched orbital angular momentum, where these terms can be directly linked to different interaction mechanisms, and it is generally true that $|J|\gg |V| \gg |D| $. The same is not true for rare earth ions, where the large spin-orbit interaction means that even interactions that are isotropic between the real electron spins on the two ions (such as Heisenberg exchange) become anisotropic in this reduced pseudo-spin Hamiltonian \citeit{wolf71, baker71}. For rare earth ions, all terms in the above equation can be a similar size, although the values can be constrained by symmetry arguments. 

Of the four interaction mechanisms introduced above, only magnetic multipole and exchange contribute to the interactions of Er\tplus pairs to first order, since the Kramers degeneracy of Er\tplus implies a first order insensitivity to electrically-mediated interactions. The magnetic multipole interaction is dominated by the magnetic dipole-dipole term \citeit{guillot-noel04}, and this can be accurately determined once the Zeeman tensors of participating ions are known:
\begin{equation}
    \bm{J^{mn}_{dd}} = \frac{\mu_0 h}{4\pi |\bm{r}|^3}\left[\bm{M^{1m} M^{2n}}-3\bm{(M^{1m}\hat{r})(M^{2n}\hat{r})^T)}\right]
\end{equation}
where $\bm{r}$ is the position vector joining the two ions. For the general case of anisotropic Zeeman tensors, $\bm{J^{mn}_{dd}}$ contributes to both the isotropic and symmetric traceless parts of Eq. \eqref{eq:shape}. 

 The direct exchange interaction is typically small and superexchange (wavefunction overlap via intervening ions) dominates because the spatial extent of $4f$ electronic wavefunctions is much smaller than typical inter-atomic spacings. Since the  environment of the sites studied here is not known, superexchange contributions cannot be approximated, but isotropic and anisotropic superexchange interactions of 1-100~GHz   between rare earth ion pairs have been seen in other materials \citeit{laplane16high, guillot-noel00,clemens83}, making this interaction similar in size to the magnetic dipole term.

\section{Experimental method}
The method for obtaining spectra of individual Er ions in silicon has been described in detail elsewhere \citeit{yin13}.Briefly, the devices used are finFETs implanted with Er in the channel region, as shown in Fig. \ref{fig:concept}.  A quantum dot is formed by the confinement provided by the corner effect and residual barriers in the access regions between the source/drain and the channel. This configuration works as a single electron transistor, and is sensitive to any nearby electrochemical potential fluctuation. If individual Er ions are resonantly optically excited into the \ere state,  they can decay non-radiatively by ionizing a nearby trap, changing the charge environment. Then, the tunneling current through the quantum dot shows a strong signal when the laser frequency matches an Er \erg-\ere transition. 

Two devices were studied here, labelled 1 and 2, with the same channel height of \SI{60}{\nano\metre}. Both were implanted with \iso{170}Er\tplus and oxygen with a ratio of 1:6, to a simulated Er\tplus concentration of $1.5\times 10^{17}$~cm$^{-3}$, and annealed at \SI{700}{\celsius} in N$_2$ for 10 minutes. Device 1 had dimensions of \num{890}$\times$\SI{100}{\nano\metre} with 780 erbium ions estimated to be in the channel. Device 2 had a dimensions of  \num{100}$\times$\SI{35}{\nano\metre}, with 30 erbium ions estimated to be in the channel.

The experiments were carried out at \SI{4.2}{\kelvin} in an Oxford helium bath cryostat with a 12~T superconducting magnet. The device was mounted on a sample rod installed with a home-built vector magnet and a single-mode fiber that was aligned with the channel region under a microscope. The beam diameter was approximately 500~$\mu$m and both laser and magnetic field were parallel to [001] and perpendicular to the plane of the device. The incident polarization was unknown, although it could be manipulated from outside the cryostat with fiber polarization paddles.

The device was biased near the Coulomb blockade region, so that discrete Coulomb peaks were visible in a current -- gate voltage trace. The gate voltage was chosen to be on the rising edge of a discrete Coulomb peak where the sensitivity to charge is the highest. Then, spectra were recorded by monitoring the current while a wavelength-tunable laser was used to excite the \erg-\ere transition of single erbium ions.

\section{Pair site spectra}\label{sec:pairspectra}
In any one device studied as described above, multiple widely separated spectral lines are observed over a range from 1529 to 1546 nm \cite{deboo20}. When the laser is resonant with one of these lines, the electrical signal shows a two-level behavior, confirming the signal arises from the excitation of a single Er\tplus ion \citeit{yin13}. The lines can typically be attributed to transitions of individual Er\tplus ions in the channel from the lowest crystal field level of \erg to the \ere multiplet. These single sites characteristically display a single line in zero field that splits in an applied field into two or four lines, the Zeeman splitting expected for sites of lower than cubic symmetry (no cubic sites have been observed optically in Si). The only exception is when the site is occupied by an Er ion with nuclear spin, in which case multiple, GHz-spaced lines are seen in both zero and applied field, commonly in groups of eight ($I=7/2$ for \iso{167}Er). Such sites are not expected in these \iso{170}Er-implanted samples.

In other cases we observe multiple lines in zero field separated by up to tens of GHz. For these lines, it is important to distinguish structure due to interacting pairs (or multi-sites) from two other cases: multiple, unrelated single Er\tplus ions with similar frequencies; and single Er\tplus ions with incidental near-degenerate crystal field ground or excited states. Pair sites can be distinguished from unrelated single sites by: a) the existence of relationships between the properties of individual lines in the spectrum (such as similar polarization dependence or Zeeman shifts) and b) by the presence of either bending of lines or anti-crossings between different lines at certain magnetic field values. In particular, almost all spin interactions will lead to some bending near zero field due to quenching of the magnetic moment of the interacting states.  Er\tplus with a near-degenerate pair of crystal field levels will also show bending or anticrossings, but only at one field value. The number of related lines seen can help further distinguish a pair site from this case, which can show up to eight lines only. 

We have observed multiple different sites with signatures of pair interactions, and discuss two examples below that represent the set. We present linear Zeeman spectra for fields along the  $z$ [001] axis, and for the first site, along the $x$ [110] axis.

\subsection{Site A at 1538.306 nm}
\begin{figure}[htbp]
	\centering
	   	\subfloat{\includegraphics[width=1.0\columnwidth]{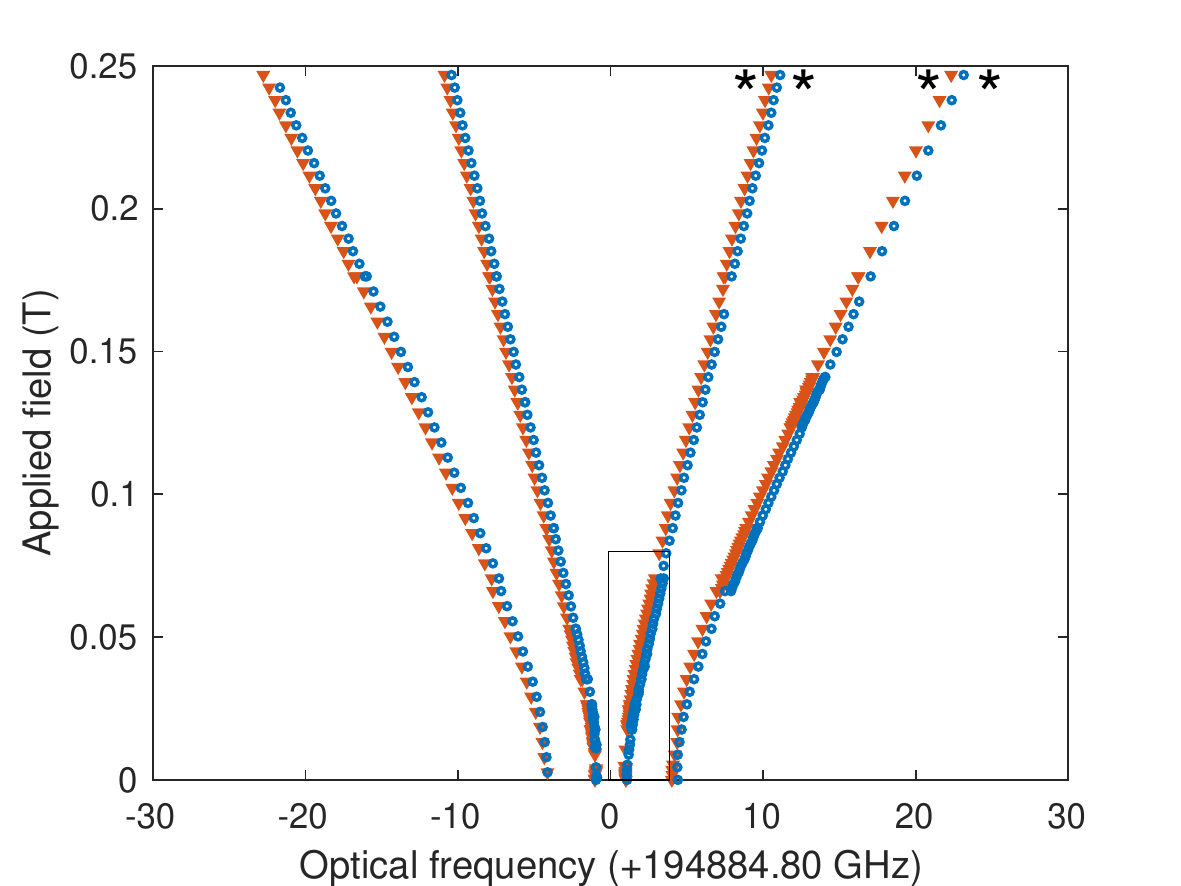}}
	   	\newline
	   	\subfloat{\includegraphics[width=0.45\columnwidth]{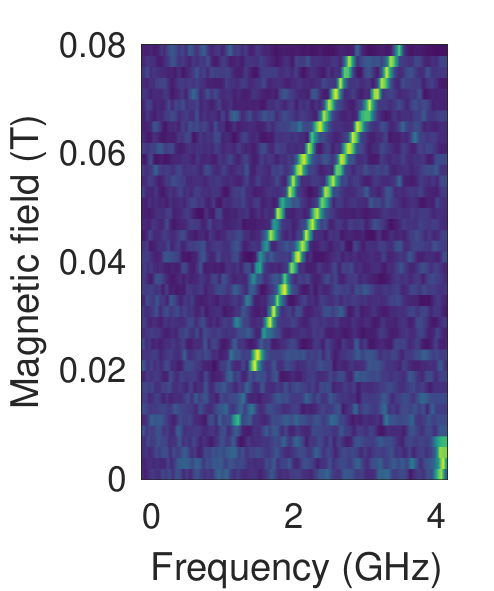}}
	   	\subfloat{\includegraphics[width=0.45\columnwidth]{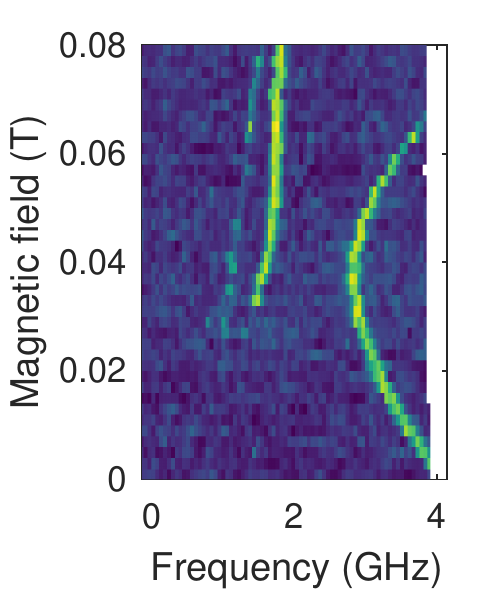}}
	\caption{\label{fig:specA} Site A, which occurs at \SI{194.88480}{\tera\hertz} (\SI{1538.306}{\nano\metre}). This spectrum consists of eight lines appearing as four slightly split pairs separated by several GHz in zero field. a) Peak locations (highest detected current) of these eight lines for fields along the lab $z$ axis, a [001] direction. Color is used to distinguish the lines of a pair. Asterisks denote lines that persist in high field, indicating those lines originate in the lowest energy ground state. The black box denotes the region for which raw spectra are shown in b), c). b) Example raw spectrum for a magnetic field along [001]. Color axis is detected current.  c) Example raw spectrum for a magnetic field along [110]. }
\end{figure}
 Fig. \ref{fig:specA}(a),  shows the lines associated with the first site, site A (seen in device 1). As described below, we attribute this site to a pair of  Er\tplus ions in non-identical sites. The spectrum displays eight lines grouped as four barely split pairs for fields along $z$. The splittings vary from 0.5 to 0.8 GHz depending on the line pair, and are roughly constant for fields above 0.05 T.  There is no Zeeman effect to first order at zero field, leading to the characteristic bending expected of a coupled site as described above.  The pair line splittings approach zero in zero field, and the two inner pairs of lines disappear (see Fig. \ref{fig:specA}(c)). When the field is rotated away from [001], the splitting within the pair increases and anticrossings appear between different pairs of lines, Fig. \ref{fig:specA}(c),  indicating that the site is anisotropic. Along $x$ ([110]), the anticrossing seen at 0.04~T in Fig. \ref{fig:specA}(b) becomes a near-crossing. The behavior seen in the $x$ and $z$ directions suggest that [110] and [001] represent symmetry axes of the site.
 
 Given that the site is anisotropic, fitting a spin Hamiltonian of the form  Eq. \ref{eq:hamil} would require spectra along several directions, which we did not collect. However, much can be determined from the spectra in Fig. \ref{fig:specA}. Eight lines in both low and high field suggest a pair site consisting of an Er\tplus ion coupled to a non-optically-active spin-half (or pseudo-spin-half)  defect. Such a system has four levels in each of the ground and excited state, and eight lines are expected when the mixing of the states is such that the optical field cannot drive flips of the non-optically active spin.  Based on the large zero field splitting, that defect must have an electronic (rather than nuclear) moment.
 
 Because four of the eight lines persist in high field, indicating that they involve transitions from the lowest energy ground state, the energy level structure must also consist of pairs of barely split levels for [001] fields. Thus, the Zeeman splitting of the optically active site generates the large splitting with $g_z = 136 \pm 4$~GHz/T in the ground state, and $g_z = 48\pm 4$~GHz/T in the excited state. Meanwhile, the optically inactive site has $g_z< 4$~GHz/T with a much larger $g_x>50~$~GHz/T in the $x$ direction. Since the optically inactive site is anisotropic with relatively large $g$ along certain directions, it can be identified as another Er ion in a different crystallographic site whose optical transitions are at a different frequency, although it may not be optically active at all.
 
Some conclusions can also be drawn about the interaction between spins. It is certainly not primarily an isotropic exchange interaction, because that is inconsistent with the zero field structure. Both symmetric and antisymmetric interactions are possible, and specifically the magnitude of the zero field splitting is consistent with a dipole-dipole interaction between two Er ions separated by $\order{4}$~\AA.

\subsection{Site B at 1534.64 nm}
\begin{figure*}[htbp]
	\centering
	   	\includegraphics[width=1.0\textwidth]{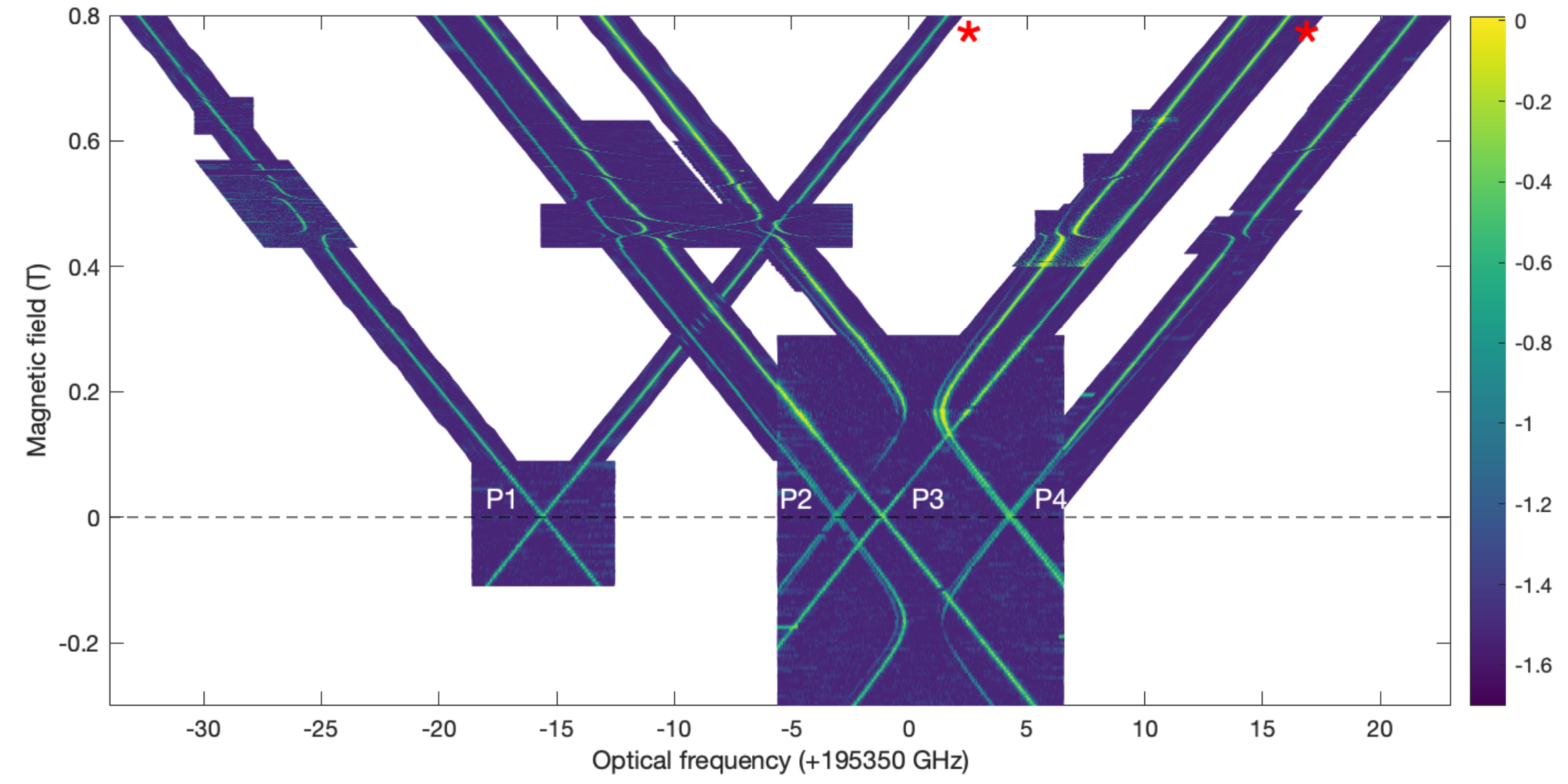}
	\caption[specB]{Composite optical spectrum for Site B at 1534.64 nm. This spectrum is composed of 23 individual spectra, either low-resolution line-following spectra or high-resolution grids around line crossings and anticrossings. The color axis is detected current, and the scale is logarithmic. All spectra have been background-subtracted. The dashed line marks zero magnetic field and the red asterisks mark the two lines still visible in fields above 2~T. The four pairs of zero field peaks are labelled P1-4. P1 and P3 split into two lines in an applied field, and P2 and P4 into four.  }
	\label{fig:specB}
\end{figure*}
Site B (device 2), shown in Fig. \ref{fig:specB}, has many more lines than site A. At zero field, four groups of lines are present, labelled peaks 1 to 4. Of these, peaks 2 and 4 are slightly split (around 150 MHz), and all peaks split near-identically in fields along the $z$ axis ([001]) at a rate of 21.82 GHz/T. The positive-frequency branches of peaks 1 and 3 persist in high fields, indicating they arise from the lowest energy ground state. Peaks 2 and 4 anti-cross near 0.2 T. The spectrum also displays many other avoided crossings between the visible lines and other, invisible lines that only gain transition intensity at the avoided crossings. The number and position of the avoided crossings establishes that there are four straight lines crossing over this spectrum from each side, with zero field offsets of the order of $\pm 100$~GHz and slopes of the order of 200 GHz/T. In total, twenty lines can be directly inferred from this spectrum. 

\begin{figure}[htbp]
	\centering
	   	\includegraphics[width=1.0\columnwidth]{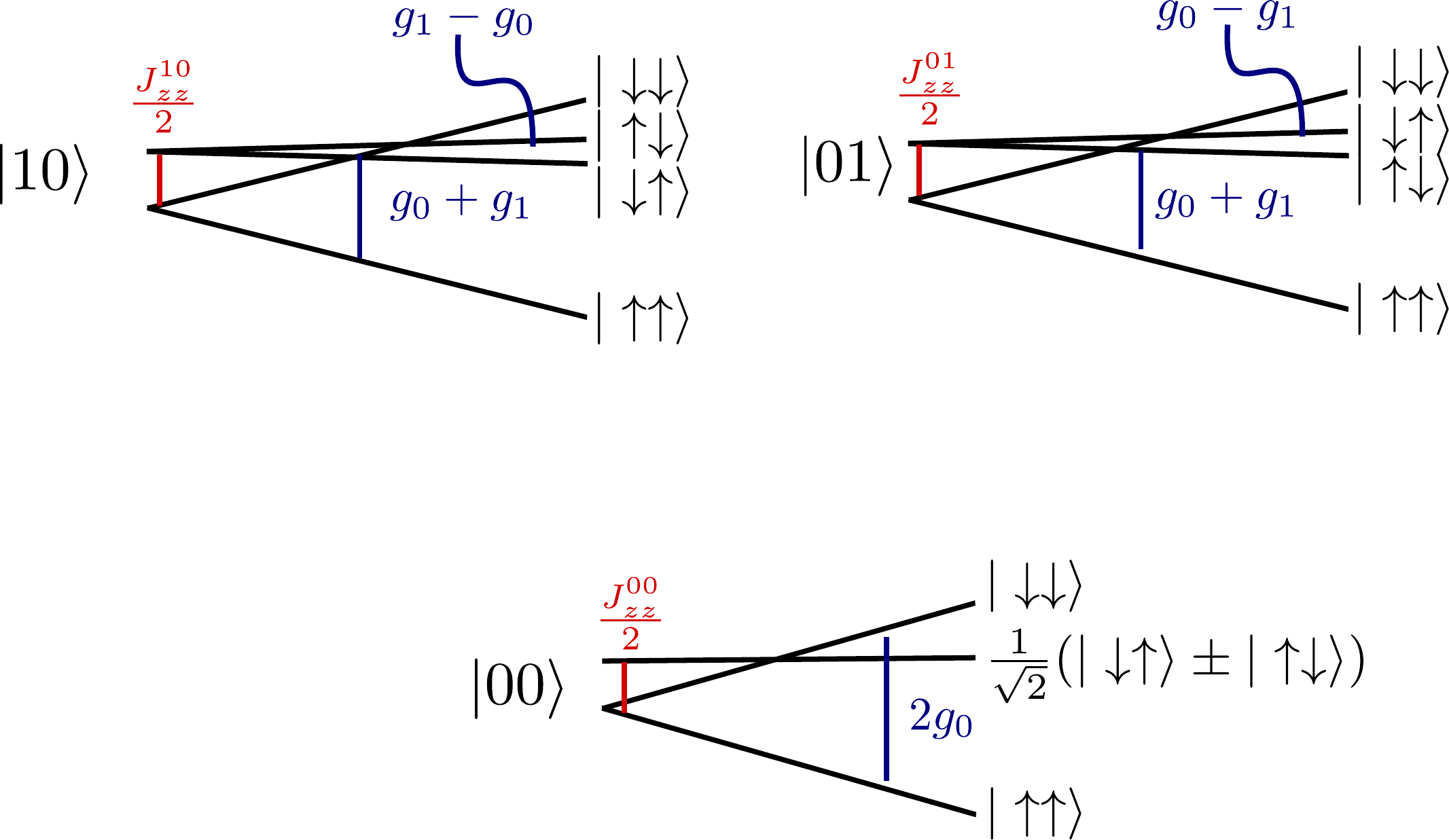}
	\caption{Prototypical energy level structure for two Er ions coupled purely via Ising spin interactions between spins in the ground and excited states, the suggested system for site B. Optical states are labelled on the left side of the energy structure, and spin eigenstates of the pair system on the right. Spin zero field and Zeeman splittings are to scale for site B. Blue arrows and text give the Zeeman splittings of the doublets in terms of ground and excited state single ion g-values in the applied field direction, and red lines and text give the optical splittings  in terms of the Ising interaction $J_{zz}$. Transitions $\ket{00}\rightarrow \ket{10}$ give rise to P1 and P2 in Fig. \ref{fig:specB}, and transitions $\ket{00}\rightarrow \ket{01}$ to P3 and P4. }
	\label{fig:levels}
\end{figure}
The spectrum satisfies both our criteria above for a pair site: lines with similar properties, and the presence of anticrossings between lines. However, in this case, we do not see the zero field bending described above. As mentioned, this structure is almost always seen for a pair site. The only pair interaction shape which can explain the lack of zero field bending is a near-Ising interaction Hamiltonian, i.e. only one term of $\mathbf{J}$ in each state is non-zero. Further, the fact that we can identify more than sixteen lines means that both ions were being excited here, so they must have similar optical frequencies and, thus, occupy identical sites. Overall, then, the spectrum can be described by two optically active Er ions occupying near-identical crystallographic sites coupled primarily by an Ising spin-Hamiltonian $\mathbf{J} = J_{zz}$ in each electronic state. Additionally, $z$ is a principal axis of the Zeeman tensor in all states and thus a symmetry axis of the site. Below, we explain how this structure leads to the spectrum seen.

The energy level structure for two ions coupled only by Ising interactions is shown in Fig. \ref{fig:levels}: it consists of two doublets in zero field for each of the three relevant electronic states (the $\ket{00}$ ground state, and the $\ket{10}$ and $\ket{01}$ singly-excited states), split by $J^{mn}_{zz}/2$. The levels are labelled according to the spin eigenstates, and the Zeeman splittings are shown as a function of the ground and excited state $g$ values of the individual ions. The spectrum requires that optical transitions in this system that preserve  spin orientation (such as $\ket{00}\ket{\uparrow \uparrow}\rightarrow \ket{10}\ket{\uparrow \uparrow}$) are allowed, and these transitions have identical Zeeman splitting of $\pm(g_0-g_1) = 21.82$~GHz for this site. These transitions explain the visible lines in Fig. \ref{fig:levels}, with peaks 1 and 2 arising from spin-preserving excitations to $\ket{10}$, and peaks 3 and 4 from excitations to $\ket{01}$. The lines appearing only at anticrossings with these lines are due to non-spin-preserving transitions. 

This model of the energy level structure is sufficient to determine ground and excited state $g$ values in the direction of the applied field ($z$) and $J_{zz}$ for each state. We find near maximal $g$ value along the $z$ direction in both states, $g_z^0 = 232$~GHz/T and $g_z^1 = 188$~GHz/T, and Ising interaction coefficients $J_{zz}^{00} = 209$~GHz, $J_{zz}^{01} = 220$~GHz, $J_{zz}^{10} = 233$~GHz. An additional optical offset between the $\ket{01}$ and $\ket{10}$ transitions of 10.8~GHz is required to explain the offset of the centroids of the peak pairs 1 and 2, and 3 and 4. As described in Section \ref{sec:pairint}, this could be caused by several interactions that we cannot distinguish in this single laser experiment: differing strain shifts due to the two ions sitting in slightly different environments, the strain shift due to the other ion of the pair, or a shift due to a diagonal optical interaction between the pair.

The Ising model is a good first approximation to the overall structure of the spectrum, but the details show other, smaller contributions must be included. First, the spin anti-crossings visible between 0.45 and 0.65~T in Fig. \ref{fig:specB} (corresponding to anti-crossings of the two doublets of a single electronic state, see Fig. \ref{fig:levels}) are crossings in an Ising model (the anticrossing at 0.15~T is an optical anticrossing, discussed later).  Second, the small (0.15~GHz) zero field splitting of the four lines contributing to peaks 2 and 4 (corresponding to transitions for peak 2, for example, of $\ket{00}\frac{1}{\sqrt{2}}\qty(\ket{\uparrow\downarrow}\pm \ket{\downarrow\uparrow})\rightarrow \ket{10}\ket{\uparrow\downarrow}, \ket{10}\ket{\downarrow\uparrow}$) is not reproduced. Both these features indicate that there is a small non-Ising contribution to the interaction spin Hamiltonian. These contributions can be isolated to a few terms in $\mathbf{J}$ in the three electronic states, because each term in $\mathbf{J}$ has a different effect on the spectrum. The lack of any zero-field anticrossings (bending) indicates that $J_{xx}\approx J_{yy} \approx 0$ for all states. This also confirms $g_x\approx g_y\approx 0$ as any transverse $g$ gives rise to $J_{xx}$ and $J_{yy}$ terms due to the magnetic dipole interaction.  The spin anticrossings at non-zero fields are caused by approximately 1~GHz $xz$ or $yz$ contributions in each state, while the zero field splitting on 2 and 4 indicates  an $xy$  contribution to $J^{00}$ only of approximately 0.15~GHz.  Additionally, there is a slight ($<2$~GHz/T) symmetric difference in the Zeeman splitting rates of the four transitions contributing to peaks 2 and 4, and a very mild ($0.1$~GHz/T) asymmetry between positive and negative Zeeman arms. These features suggest the two Er ions are not exactly identical, with slightly different $g$ values or orientations of the $g$-tensor. Consequently, the field-independent  ground state doublet gains a slight splitting in a magnetic field, with the eigenfunctions better described with new eigenstates $\ket{\downarrow \uparrow}$ and $\ket{\uparrow\downarrow }$.  These differences likely indicate slightly different crystalline environments for the two ions, although canting of the moments due to the ion-ion interaction \cite{erdos66} could also explain small differences in orientation. 


\begin{figure}[htbp]
	\centering
	   	\includegraphics[width=1.0\columnwidth]{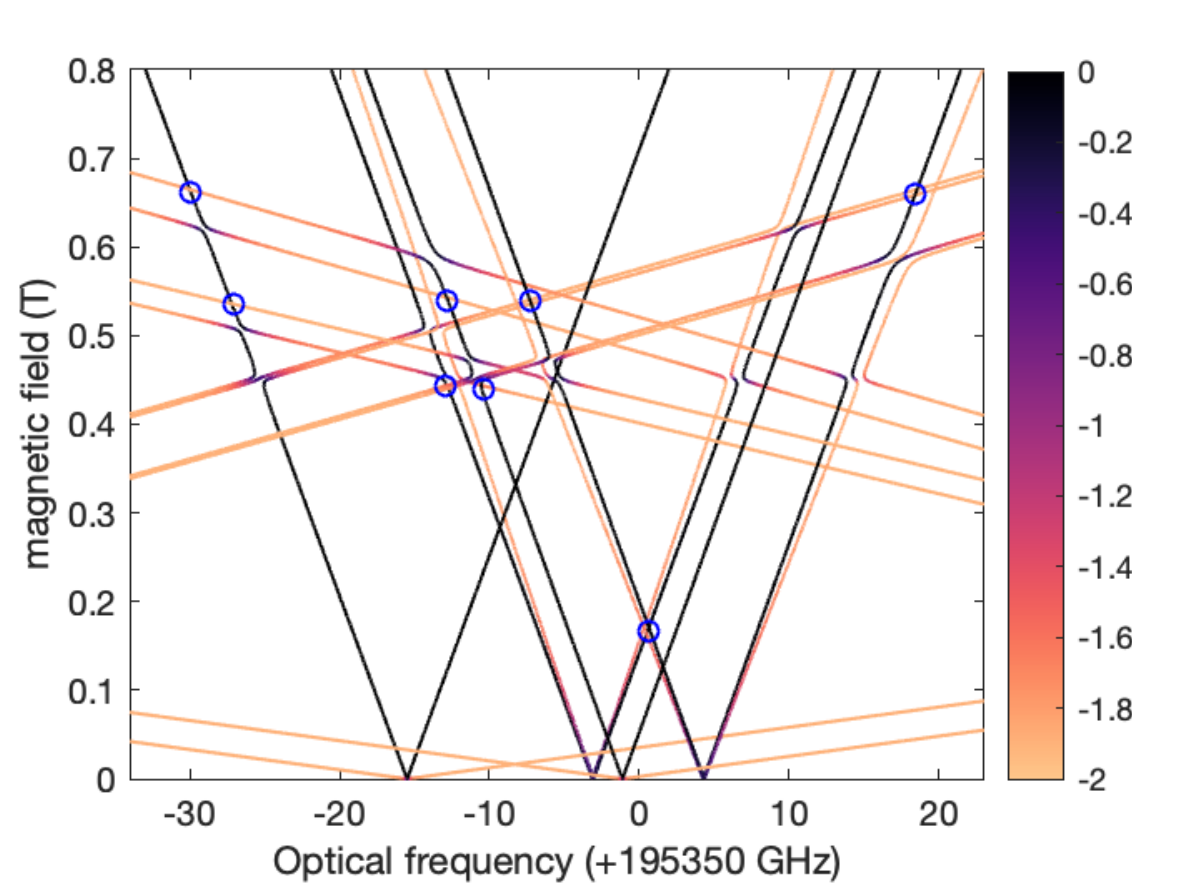}
	\caption{Simulation of the optical spectrum of Fig. \ref{fig:specB} using a  near-Ising spin Hamiltonian model for each of the three electronic states, as described in the text. Color indicates the transition probability assuming a spin-preserving selection rule. The color axis is logarithmic and matches Fig. \ref{fig:specB}, $\log_{10}(P+0.03)$ where $P$ is the spin wavefunction overlap defined in-text. Blue circles mark the position of optical anti-crossings seen in the data.}
	\label{fig:sim}
\end{figure}
In Fig. \ref{fig:sim}, we show that including these extra terms in the spin Ising model reproduces the key elements of Fig. \ref{fig:specB} quite well. In the figure, we simulate the optical spectrum expected from near-Ising spin Hamiltonians for $\ket{00}$, $\ket{10}$, and $\ket{01}$ with $g_z$ and $J_{zz}$ values as given above, $g_x=g_y=0$ in all states, and with additional terms as follows: for ion 2, $g_z$ decreased by 1.5 GHz in the ground state and 1 GHz in the excited state (to account for the different Zeeman splittings of the lines in P2 and P4); and the addition of an antisymmetric term (Eq. \eqref{eq:shape}) $\mathbf{D}^{00} = [2, 0, 0.2]$~GHz and $\mathbf{D}^{10} = \mathbf{D}^{01} = [1, 0, 0]$~GHz to account for the additional $J_{xy}$, $J_{xz}$ and J$_{yz}$ terms as described above. In the figure, energy differences were calculated between ground and excited state spin Hamiltonians with optical frequency shifts chosen to match the experimental spectrum. Finally, optical transition intensities (color axis) were estimated assuming a spin-preserving selection rule, by calculating the spin wavefunction overlap $P = |\braket{\Psi_f}{\Psi_i}|^2$ where $\ket{\Psi_i}$ and $\ket{\Psi_f}$ are the spin wavefunctions of the states determined from the appropriate ground and excited spin Hamiltonians. The model reproduces all spin anticrossings seen in the spectrum, however, there are several optical anticrossings that are not reproduced in this spin Hamiltonian simulation, most notably the large anticrossing at 0.15~T. These indicate an additional small optical interaction, discussed in Section \ref{sec:opticalanti} below.

Given the spin model described above can well explain the  spectrum observed, we now expand on the physical coupled site that could lead to the spin parameters determined. The two constituent Er ions have identical g-tensors with one principal axis along [001] in both sites, which requires a site symmetry of at least C$_2$. The g-tensors are near maximal and approximately Ising-like ($g_\perp\approx 0$), implying the ground and excited states are near-pure $\ket{M_J = \pm 15/2}$ and $\ket{M_J = \pm 13/2}$ doublets. This shape commonly occurs in rare earth Kramers ions in crystals  for certain site symmetries or certain values of crystal field parameters  (e.g. \cite{carey09, wolf00}). Indeed, an Er site with an Ising-like g-tensor with similar $g_{||}$ in both ground and excited states has been previously observed in silicon \cite{vinh03}, suggested to be Er in either a substitutional or interstitial $T_d$ site  ( although the latter is strongly preferred for Er \citeit{kenyon05}) distorted down to $C_{2v}$. 

That site had a different optical frequency (1538 nm) and was oriented differently to our coupled site, with $g_{||}$ along [110]. However, it is likely our coupled site is related, consisting of two Er ions sitting in identical $T_d$ sites. These individual sites may be distorted to lower symmetry by the incorporation of individual Er (possibly coordinated with oxygen or another impurity), or the presence of the other Er may lower the symmetry. For instance, placing Er ions in the two interstitial  $T_d$ sites separated along [001] by the unit cell spacing 5.4~\AA\ would lower the symmetry to $C_{2v}$ with the $C_2$ axis along [001], consistent with our measured g-tensors. The resulting pair site would be symmetric about the intervening substitutional T$_d$ site, whose symmetry would be lowered to D$_{2d}$. It is likely each Er ion would be displaced from the T$_d$ site by the strain caused by the other Er ion (spacing different to 5.4~\AA), but this displacement must be along the [001] to maintain the D$_{2d}$ symmetry of the intervening site. Such a displacement would lower the Er site symmetry to C$_2$, which is still consistent with the g-tensors presented here.

Such a description is also consistent with the optical selection rules. For our geometry, with the light propagating along the C$_2$ axis, only $\sigma$ polarization is accessible for both electric and magnetic dipole components of the light. Both these have the same $\Delta M_J = \pm 1$ selection rule, which allows only spin preserving transitions, such as  $ \ket{00}\ket{\frac{-15}{2} \frac{-15}{2}}\rightarrow \ket{10}\ket{\frac{-13}{2} \frac{-15}{2}}$ (equivalent to $\ket{00}\ket{\downarrow\downarrow}\rightarrow \ket{10}\ket{\downarrow\downarrow}$).

Finally, we consider the origin of the spin-spin interaction. For a site with Ising g-tensors, an overall $C_2$ or higher symmetry of the individual sites, and $D_{2d}$ symmetry of the whole site, both magnetic dipole-dipole interactions and exchange interactions result in an Ising interaction Hamiltonian, so these interactions cannot be separated on the basis of interaction shape. This is unfortunate, because knowledge of the magnetic dipole component would allow the distance between the ions to be determined, since the magnetic dipole moments are known. We can, however, make some estimates of the relative dipole and exchange components. The interaction is not entirely dipole-dipole, because that would imply that $J_{zz}$ is proportional to $g_z$ for each electronic state and thus a larger interaction would be expected in the ground state, instead of in the excited states as seen here. The smallest possible exchange contribution that can explain the $J_{zz}$ values seen is 20\% (for two ions separated by 3.25\AA) but a much larger exchange contribution is required for larger separations. For instance, at the 5.4~\AA\ lattice spacing of the tetrahedral sites, $>80\%$ of $J_{zz}$ must come from exchange. Such a contribution is reasonable: in Kramers rare earth ion pair sites, exchange interactions have been observed to be similar in size to magnetic dipole-dipole out to separations of 9~\AA \cite{birgeneau68, brower66, guillot-noel04}.

\subsubsection{Optical anti-crossing}\label{sec:opticalanti}
The large (1.5~GHz) anti-crossing at 0.15~T  in Fig. \ref{fig:specB}, and several smaller anticrossings at higher field are not reproduced in the spin Hamiltonian simulation of Fig. \ref{fig:sim}, indicating that these are due to optical interactions: they arise from anticrossings between spin states belonging to different optical excited states $\ket{10}$ and $\ket{01}$. The large anticrossing at 0.15~T, in particular, has two interesting features: first, one of the two anti-crossing states has near zero optical transition strength, indicating the transition to this state has become forbidden; second, the position of this dark state flips when the direction of the magnetic field is reversed, being on the low frequency side of the anti-crossing for positive fields and the high frequency side for negative fields. Here, we describe this behavior with a simplified model. 

From the spin model, we can determine that this large anti-crossing occurs between the $\ket{10}\ket{\downarrow\uparrow}$ and $\ket{01}\ket{\downarrow\uparrow}$ states. The behavior can be understood by considering a new reduced Hamiltonian acting on the four electronic states with spin component $\ket{\downarrow\uparrow}$, which we will write as $\ket{0'0'},\ket{1'0'},\ket{0'1'}, \ket{1'1'}$. Then, the Hamiltonian of the system can be generically represented as:
\begin{equation}
    H = H_1+H_2+H_{12}
\end{equation}
where $H_1, H_2$ represent the Hamiltonians of the individual ions, and $H_{12}$ describes the interaction. When the two ions are resonant at the anticrossing, the wavefunctions are the three symmetric states $\ket{0'0'}, \ket{1'1'}$ and $\frac{1}{\sqrt{2}}\left[\ket{0'1'}+\ket{1'0'}\right]$ and the antisymmetric state $\frac{1}{\sqrt{2}}\left[\ket{0'1'}-\ket{1'0'}\right]$. The latter two states are split by $2\bra{0'1'}H_{12}\ket{1'0'} = 1.6$~GHz. Transitions between the symmetric states retain the same selection rules for the individual sites, so are allowed in our $\sigma$ polarization. However, transitions from $\ket{0'0'}$ to the antisymmetric state are forbidden:  the  transition operators (electric dipole and magnetic dipole) are identical for the two single-ion transitions for $\sigma$ light due to the overall D$_{2d}$ symmetry of the coupled site, and so the contributions from the two component states to the antisymmetric wavefunction cancel. Thus, one branch of the anti-crossing is dark. Note that the transition can be allowed for $\pi$ light. An example is seen in the ensemble coupled site in Pr:LaF$_3$ studied by Ref. \citeit{lukac89}). 

The fact that the dark state flips across the anticrossing when the field is reversed implies that the interaction is  purely or mostly odd under time reversal symmetry: for negative fields, the relevant anticrossing states have spin projection $\ket{\uparrow\downarrow}$, the time-reversed copy of the positive magnetic field states $\ket{\downarrow\uparrow}$. Interaction mechanisms such as exchange are even under time reversal; the odd symmetry means the interaction must contain a magnetic term.  A possible interaction is a photon-mediated (Forster) interaction with a magnetic dipole--electric dipole character\note{References?}. The optical transition of Er\tplus is frequently of mixed electric/magnetic dipole character, making this interaction feasible. However, since this electrical detection method does not provide information on the optical transition dipole moments, it is difficult to estimate the size of this odd interaction in Si.  \note{Alexey, can you check over this and add any extra info?}

\section{Discussion}
Over several Er implanted devices with similar Er concentrations to those studied here, we have observed about 200 unique resonances. Of these, approximately 10\% show signatures of pairing according to our definition in Section \ref{sec:pairspectra}. This high proportion is, perhaps, not unexpected as Er is well known to cluster in silicon \citeit{kenyon05}. Studying these pair sites will give more information about the sorts of clusters formed. Here, we chose two representative pair sites with relatively simple optical structure to investigate in more detail.

Site B showed both optical and spin interactions. We saw a 1.5~GHz off-diagonal optical interaction with an unusual odd time-reversal symmetry, and cannot rule out a diagonal contribution to the 10.8~GHz optical frequency detuning of the two ions of the pair. These interactions are similar to those seen in several other rare earth crystals, including Pr:LaF$_3$ \citeit{lukac89}, GdCl$_3$ and Gd(OH)$_3$ \citeit{cone73}, although much larger optical interactions are possible \citeit{cone75}.

Both sites showed strong spin-spin coupling, 10-200~GHz. In each case, the isotropic exchange contribution was unmeasurable, $<0.1$~GHz, and both magnetic dipole and exchange were likely to contribute to the coupling. In particular, for site B, we showed that at least 25\%, 50~GHz, of the spin-spin interaction must be an anisotropic exchange term, with the remainder arising from the magnetic dipole-dipole contribution. The exchange contribution differed significantly between the ground and excited states, which is consistent with previous measurements of spin interactions in photoexcited ensembles, where superexchange contributions varied by up to an order of magnitude between levels \citeit{clemens83, prinz66}.

The spin interactions seen here can be compared with previous measurements of isotropic exchange, anisotropic exchange, and magnetic dipole spin interactions for different rare earth pairs in ensembles, reviewed by Guillot-No\"el et al. \citeit{guillot-noel04}.   Our magnetic dipole interactions are much larger than most other materials, because the two ions have nearly the maximum $g_z$ possible for a rare earth ion, and interact along the direction of strongest coupling. The large anisotropic exchange interaction is more surprising. It is nearly 2 orders of magnitude greater than most of the materials listed by  Guillot-No\"el et al., even those with small pair spacings for rare earth ions of 3-4 \AA. However, it is exactly pairs with strong interactions that are difficult to measure using the techniques employed in most of those studies. Most measurements used EPR, with which it is challenging to even observe lines spaced by $\approx100$~GHz in zero field. The situation is slightly improved when using optical methods, since they have a much larger bandwidth. However, for large interactions comparable to the strain shifts of the pair sites, it would be difficult to distinguish these two contributions to the satellite structure and associate lines with particular pair sites. Thus, the very strong exchange interactions seen here may be more common in rare earth pairs than previous ensemble pair site measurements suggest. Similar contributions from anisotropic exchange have been seen, for instance, using optical measurements in  NdCl$_3$ (ion separation of 4.2~\AA)\citeit{prinz66}, and studying magnetic ordering in GdVO$_4$ (ion separation of 3.9\AA) \citeit{abraham92}.

We showed that site B was best explained by an Ising spin interaction, but that this was perturbed slightly with very small non-Ising terms, 0.1-1\%. These contributions are likely due to strain in the site, causing a slight deviation from the overall proposed D$_{2d}$ symmetry. We were able to measure these very small contributions remarkably accurately because we studied only a single instance of the site; in an ensemble, this type of strain variation leads to inhomogeneous broadening, and so only average effects of strain can be characterized. Individually studying multiple instances of the same site would be an excellent way to better understand the microscopic nature of strain effects in rare earth crystals. Among the 20 pair sites we have seen so far, site B has not reoccurred, although we have seen a related site with an Ising coupling and similar g values, so a wider survey of sites would be required. We also have not seen evidence of an Er\tplus single site with the C$_{2v}$ symmetry and Ising $g$-tensors of the two constituent sites of this pair site. This does not necessarily imply that the site cannot be occupied by only one Er\tplus ion. There is no requirement for the single site to be spectrally close to the pair since strain shifts of pair sites can be well in excess of 500~GHz (4~nm) \citeit{yamaguchi98}, and we have not yet characterized the symmetry of all 200-odd single Er\tplus sites observed in these devices.

Proposals for quantum computing with rare earth ion arrays often consider using the optical excitation blockade interaction to perform two qubit gates \citeit{ohlsson02, ahlefeldt20}, which is the frequency shift of one ion on the excitation of another. In our coupled model Fig. \ref{fig:concept}, this corresponds to the frequency difference between, for instance, the spin-preserving transitions "a" ($\ket{00}\rightarrow \ket{10}$) and "c" ($\ket{10}\rightarrow \ket{11}$). Here, a blockade interaction of approximately 30~GHz arises from the spin interactions alone. This is three orders of magnitude greater than the electric dipole blockade interaction seen for nearest-neighbor Eu\tplus pairs \citeit{ahlefeldt13precision}, and six orders of magnitude larger than the interaction used for the only rare earth two qubit gate to date \citeit{longdell04a}. This type of strong magnetic dipole interaction offers the prospect of fast gates times for blockade \citeit{ohlsson02} and direct interaction gates \citeit{kinos21,grimm21}. 

We do note that a first step to using these interactions for quantum computing in silicon, specifically, is to demonstrate initialization and spin-state selective readout of individual ions, which has not been possible so far with this electro-optical detection method. However, recent measurements have shown that implanted Er\tplus in Si does have good spin coherence properties \citeit{hughes21}, and all-optical methods for detection of single Er\tplus ions are being developed \citeit{berkman21,weiss21}.

\section{Conclusion}
We presented optical measurements of two different examples of individual Er\tplus pair sites in silicon: pair in identical crystallographic sites and a second pair in non-identical sites. We showed that the identical pair displayed Ising-like $g$-tensors and a very strong, 200~GHz, Ising spin interaction in addition to a 1.5~GHz optical interaction between the ions of the pair. We were able to resolve a small distortion to the Ising symmetry of 0.1\% attributed to strain in the site. To our knowledge, this is the first study of a rare earth pair site seen either in ensemble or single instance coupled simultaneously by spin and optical interactions, and the rich structure and strong interactions seen here suggest such coupled sites could be a useful resource for quantum computing.

\section{Acknowledgements}
We thank John Bartholomew for insightful discussions, and acknowledge funding from the Australian Research Council Centre of Excellence for Quantum Computation and Communication Technology (Grant No. CE170100012).  We acknowledge the AFAiiR node of the NCRIS Heavy Ion Capability for access to ion-implantation facilities.

GH and RLA contributed equally to this work.


\begin{thebibliography}{57}%
\makeatletter
\providecommand \@ifxundefined [1]{%
 \@ifx{#1\undefined}
}%
\providecommand \@ifnum [1]{%
 \ifnum #1\expandafter \@firstoftwo
 \else \expandafter \@secondoftwo
 \fi
}%
\providecommand \@ifx [1]{%
 \ifx #1\expandafter \@firstoftwo
 \else \expandafter \@secondoftwo
 \fi
}%
\providecommand \natexlab [1]{#1}%
\providecommand \enquote  [1]{``#1''}%
\providecommand \bibnamefont  [1]{#1}%
\providecommand \bibfnamefont [1]{#1}%
\providecommand \citenamefont [1]{#1}%
\providecommand \href@noop [0]{\@secondoftwo}%
\providecommand \href [0]{\begingroup \@sanitize@url \@href}%
\providecommand \@href[1]{\@@startlink{#1}\@@href}%
\providecommand \@@href[1]{\endgroup#1\@@endlink}%
\providecommand \@sanitize@url [0]{\catcode `\\12\catcode `\$12\catcode
  `\&12\catcode `\#12\catcode `\^12\catcode `\_12\catcode `\%12\relax}%
\providecommand \@@startlink[1]{}%
\providecommand \@@endlink[0]{}%
\providecommand \url  [0]{\begingroup\@sanitize@url \@url }%
\providecommand \@url [1]{\endgroup\@href {#1}{\urlprefix }}%
\providecommand \urlprefix  [0]{URL }%
\providecommand \Eprint [0]{\href }%
\providecommand \doibase [0]{https://doi.org/}%
\providecommand \selectlanguage [0]{\@gobble}%
\providecommand \bibinfo  [0]{\@secondoftwo}%
\providecommand \bibfield  [0]{\@secondoftwo}%
\providecommand \translation [1]{[#1]}%
\providecommand \BibitemOpen [0]{}%
\providecommand \bibitemStop [0]{}%
\providecommand \bibitemNoStop [0]{.\EOS\space}%
\providecommand \EOS [0]{\spacefactor3000\relax}%
\providecommand \BibitemShut  [1]{\csname bibitem#1\endcsname}%
\let\auto@bib@innerbib\@empty
\bibitem [{\citenamefont {Awschalom}\ \emph {et~al.}(2018)\citenamefont
  {Awschalom}, \citenamefont {Hanson}, \citenamefont {Wrachtrup},\ and\
  \citenamefont {Zhou}}]{awschalom18}%
  \BibitemOpen
  \bibfield  {author} {\bibinfo {author} {\bibfnamefont {D.~D.}\ \bibnamefont
  {Awschalom}}, \bibinfo {author} {\bibfnamefont {R.}~\bibnamefont {Hanson}},
  \bibinfo {author} {\bibfnamefont {J.}~\bibnamefont {Wrachtrup}},\ and\
  \bibinfo {author} {\bibfnamefont {B.~B.}\ \bibnamefont {Zhou}},\ }\href
  {https://doi.org/10.1038/s41566-018-0232-2} {\bibfield  {journal} {\bibinfo
  {journal} {Nature Photonics}\ }\textbf {\bibinfo {volume} {12}},\ \bibinfo
  {pages} {516} (\bibinfo {year} {2018})}\BibitemShut {NoStop}%
\bibitem [{\citenamefont {Bradley}\ \emph {et~al.}(2019)\citenamefont
  {Bradley}, \citenamefont {Randall}, \citenamefont {Abobeih}, \citenamefont
  {Berrevoets}, \citenamefont {Degen}, \citenamefont {Bakker}, \citenamefont
  {Markham}, \citenamefont {Twitchen},\ and\ \citenamefont
  {Taminiau}}]{bradley19}%
  \BibitemOpen
  \bibfield  {author} {\bibinfo {author} {\bibfnamefont {C.~E.}\ \bibnamefont
  {Bradley}}, \bibinfo {author} {\bibfnamefont {J.}~\bibnamefont {Randall}},
  \bibinfo {author} {\bibfnamefont {M.~H.}\ \bibnamefont {Abobeih}}, \bibinfo
  {author} {\bibfnamefont {R.~C.}\ \bibnamefont {Berrevoets}}, \bibinfo
  {author} {\bibfnamefont {M.~J.}\ \bibnamefont {Degen}}, \bibinfo {author}
  {\bibfnamefont {M.~A.}\ \bibnamefont {Bakker}}, \bibinfo {author}
  {\bibfnamefont {M.}~\bibnamefont {Markham}}, \bibinfo {author} {\bibfnamefont
  {D.~J.}\ \bibnamefont {Twitchen}},\ and\ \bibinfo {author} {\bibfnamefont
  {T.~H.}\ \bibnamefont {Taminiau}},\ }\href
  {https://doi.org/10.1103/PhysRevX.9.031045} {\bibfield  {journal} {\bibinfo
  {journal} {Physical Review X}\ }\textbf {\bibinfo {volume} {9}},\ \bibinfo
  {pages} {031045} (\bibinfo {year} {2019})}\BibitemShut {NoStop}%
\bibitem [{\citenamefont {Kolesov}\ \emph {et~al.}(2012)\citenamefont
  {Kolesov}, \citenamefont {Xia}, \citenamefont {Reuter}, \citenamefont
  {St{\"o}hr}, \citenamefont {Zappe}, \citenamefont {Meijer}, \citenamefont
  {Hemmer},\ and\ \citenamefont {Wrachtrup}}]{kolesov12}%
  \BibitemOpen
  \bibfield  {author} {\bibinfo {author} {\bibfnamefont {R.}~\bibnamefont
  {Kolesov}}, \bibinfo {author} {\bibfnamefont {K.}~\bibnamefont {Xia}},
  \bibinfo {author} {\bibfnamefont {R.}~\bibnamefont {Reuter}}, \bibinfo
  {author} {\bibfnamefont {R.}~\bibnamefont {St{\"o}hr}}, \bibinfo {author}
  {\bibfnamefont {A.}~\bibnamefont {Zappe}}, \bibinfo {author} {\bibfnamefont
  {J.}~\bibnamefont {Meijer}}, \bibinfo {author} {\bibfnamefont {P.~R.}\
  \bibnamefont {Hemmer}},\ and\ \bibinfo {author} {\bibfnamefont
  {J.}~\bibnamefont {Wrachtrup}},\ }\href {https://doi.org/10.1038/ncomms2034}
  {\bibfield  {journal} {\bibinfo  {journal} {Nature Communications}\ }\textbf
  {\bibinfo {volume} {3}},\ \bibinfo {pages} {1029} (\bibinfo {year}
  {2012})}\BibitemShut {NoStop}%
\bibitem [{\citenamefont {Utikal}\ \emph {et~al.}(2014)\citenamefont {Utikal},
  \citenamefont {Eichhammer}, \citenamefont {Petersen}, \citenamefont {Renn},
  \citenamefont {G{\"o}tzinger},\ and\ \citenamefont {Sandoghdar}}]{utikal14}%
  \BibitemOpen
  \bibfield  {author} {\bibinfo {author} {\bibfnamefont {T.}~\bibnamefont
  {Utikal}}, \bibinfo {author} {\bibfnamefont {E.}~\bibnamefont {Eichhammer}},
  \bibinfo {author} {\bibfnamefont {L.}~\bibnamefont {Petersen}}, \bibinfo
  {author} {\bibfnamefont {A.}~\bibnamefont {Renn}}, \bibinfo {author}
  {\bibfnamefont {S.}~\bibnamefont {G{\"o}tzinger}},\ and\ \bibinfo {author}
  {\bibfnamefont {V.}~\bibnamefont {Sandoghdar}},\ }\href
  {https://doi.org/10.1038/ncomms4627} {\bibfield  {journal} {\bibinfo
  {journal} {Nature Communications}\ }\textbf {\bibinfo {volume} {5}},\
  \bibinfo {pages} {3627} (\bibinfo {year} {2014})}\BibitemShut {NoStop}%
\bibitem [{\citenamefont {Nakamura}\ \emph {et~al.}(2014)\citenamefont
  {Nakamura}, \citenamefont {Yoshihiro}, \citenamefont {Inagawa}, \citenamefont
  {Fujiyoshi},\ and\ \citenamefont {Matsushita}}]{nakamura14}%
  \BibitemOpen
  \bibfield  {author} {\bibinfo {author} {\bibfnamefont {I.}~\bibnamefont
  {Nakamura}}, \bibinfo {author} {\bibfnamefont {T.}~\bibnamefont {Yoshihiro}},
  \bibinfo {author} {\bibfnamefont {H.}~\bibnamefont {Inagawa}}, \bibinfo
  {author} {\bibfnamefont {S.}~\bibnamefont {Fujiyoshi}},\ and\ \bibinfo
  {author} {\bibfnamefont {M.}~\bibnamefont {Matsushita}},\ }\href
  {https://doi.org/10.1038/srep07364} {\bibfield  {journal} {\bibinfo
  {journal} {Scientific Reports}\ }\textbf {\bibinfo {volume} {4}},\ \bibinfo
  {pages} {7364} (\bibinfo {year} {2014})}\BibitemShut {NoStop}%
\bibitem [{\citenamefont {Kolesov}\ \emph {et~al.}(2013)\citenamefont
  {Kolesov}, \citenamefont {Xia}, \citenamefont {Reuter}, \citenamefont
  {Jamali}, \citenamefont {St{\"o}hr}, \citenamefont {Inal}, \citenamefont
  {Siyushev},\ and\ \citenamefont {Wrachtrup}}]{kolesov13}%
  \BibitemOpen
  \bibfield  {author} {\bibinfo {author} {\bibfnamefont {R.}~\bibnamefont
  {Kolesov}}, \bibinfo {author} {\bibfnamefont {K.}~\bibnamefont {Xia}},
  \bibinfo {author} {\bibfnamefont {R.}~\bibnamefont {Reuter}}, \bibinfo
  {author} {\bibfnamefont {M.}~\bibnamefont {Jamali}}, \bibinfo {author}
  {\bibfnamefont {R.}~\bibnamefont {St{\"o}hr}}, \bibinfo {author}
  {\bibfnamefont {T.}~\bibnamefont {Inal}}, \bibinfo {author} {\bibfnamefont
  {P.}~\bibnamefont {Siyushev}},\ and\ \bibinfo {author} {\bibfnamefont
  {J.}~\bibnamefont {Wrachtrup}},\ }\href
  {https://doi.org/10.1103/PhysRevLett.111.120502} {\bibfield  {journal}
  {\bibinfo  {journal} {Physical Review Letters}\ }\textbf {\bibinfo {volume}
  {111}},\ \bibinfo {pages} {120502} (\bibinfo {year} {2013})}\BibitemShut
  {NoStop}%
\bibitem [{\citenamefont {Yin}\ \emph {et~al.}(2013)\citenamefont {Yin},
  \citenamefont {Rancic}, \citenamefont {{de Boo}}, \citenamefont {Stavrias},
  \citenamefont {McCallum}, \citenamefont {Sellars},\ and\ \citenamefont
  {Rogge}}]{yin13}%
  \BibitemOpen
  \bibfield  {author} {\bibinfo {author} {\bibfnamefont {C.}~\bibnamefont
  {Yin}}, \bibinfo {author} {\bibfnamefont {M.}~\bibnamefont {Rancic}},
  \bibinfo {author} {\bibfnamefont {G.~G.}\ \bibnamefont {{de Boo}}}, \bibinfo
  {author} {\bibfnamefont {N.}~\bibnamefont {Stavrias}}, \bibinfo {author}
  {\bibfnamefont {J.~C.}\ \bibnamefont {McCallum}}, \bibinfo {author}
  {\bibfnamefont {M.~J.}\ \bibnamefont {Sellars}},\ and\ \bibinfo {author}
  {\bibfnamefont {S.}~\bibnamefont {Rogge}},\ }\href
  {https://doi.org/10.1038/nature12081} {\bibfield  {journal} {\bibinfo
  {journal} {Nature}\ }\textbf {\bibinfo {volume} {497}},\ \bibinfo {pages}
  {91} (\bibinfo {year} {2013})}\BibitemShut {NoStop}%
\bibitem [{\citenamefont {{Groot-Berning}}\ \emph {et~al.}(2019)\citenamefont
  {{Groot-Berning}}, \citenamefont {Kornher}, \citenamefont {Jacob},
  \citenamefont {Stopp}, \citenamefont {Dawkins}, \citenamefont {Kolesov},
  \citenamefont {Wrachtrup}, \citenamefont {Singer},\ and\ \citenamefont
  {{Schmidt-Kaler}}}]{groot-berning19}%
  \BibitemOpen
  \bibfield  {author} {\bibinfo {author} {\bibfnamefont {K.}~\bibnamefont
  {{Groot-Berning}}}, \bibinfo {author} {\bibfnamefont {T.}~\bibnamefont
  {Kornher}}, \bibinfo {author} {\bibfnamefont {G.}~\bibnamefont {Jacob}},
  \bibinfo {author} {\bibfnamefont {F.}~\bibnamefont {Stopp}}, \bibinfo
  {author} {\bibfnamefont {S.~T.}\ \bibnamefont {Dawkins}}, \bibinfo {author}
  {\bibfnamefont {R.}~\bibnamefont {Kolesov}}, \bibinfo {author} {\bibfnamefont
  {J.}~\bibnamefont {Wrachtrup}}, \bibinfo {author} {\bibfnamefont
  {K.}~\bibnamefont {Singer}},\ and\ \bibinfo {author} {\bibfnamefont
  {F.}~\bibnamefont {{Schmidt-Kaler}}},\ }\href
  {https://doi.org/10.1103/PhysRevLett.123.106802} {\bibfield  {journal}
  {\bibinfo  {journal} {Physical Review Letters}\ }\textbf {\bibinfo {volume}
  {123}},\ \bibinfo {pages} {106802} (\bibinfo {year} {2019})}\BibitemShut
  {NoStop}%
\bibitem [{\citenamefont {Chen}\ \emph {et~al.}(2020)\citenamefont {Chen},
  \citenamefont {Raha}, \citenamefont {Phenicie}, \citenamefont {Ourari},\ and\
  \citenamefont {Thompson}}]{chen20}%
  \BibitemOpen
  \bibfield  {author} {\bibinfo {author} {\bibfnamefont {S.}~\bibnamefont
  {Chen}}, \bibinfo {author} {\bibfnamefont {M.}~\bibnamefont {Raha}}, \bibinfo
  {author} {\bibfnamefont {C.~M.}\ \bibnamefont {Phenicie}}, \bibinfo {author}
  {\bibfnamefont {S.}~\bibnamefont {Ourari}},\ and\ \bibinfo {author}
  {\bibfnamefont {J.~D.}\ \bibnamefont {Thompson}},\ }\href
  {https://doi.org/10.1126/science.abc7821} {\bibfield  {journal} {\bibinfo
  {journal} {Science}\ }\textbf {\bibinfo {volume} {370}},\ \bibinfo {pages}
  {592} (\bibinfo {year} {2020})}\BibitemShut {NoStop}%
\bibitem [{\citenamefont {Raha}\ \emph {et~al.}(2020)\citenamefont {Raha},
  \citenamefont {Chen}, \citenamefont {Phenicie}, \citenamefont {Ourari},
  \citenamefont {Dibos},\ and\ \citenamefont {Thompson}}]{raha20}%
  \BibitemOpen
  \bibfield  {author} {\bibinfo {author} {\bibfnamefont {M.}~\bibnamefont
  {Raha}}, \bibinfo {author} {\bibfnamefont {S.}~\bibnamefont {Chen}}, \bibinfo
  {author} {\bibfnamefont {C.~M.}\ \bibnamefont {Phenicie}}, \bibinfo {author}
  {\bibfnamefont {S.}~\bibnamefont {Ourari}}, \bibinfo {author} {\bibfnamefont
  {A.~M.}\ \bibnamefont {Dibos}},\ and\ \bibinfo {author} {\bibfnamefont
  {J.~D.}\ \bibnamefont {Thompson}},\ }\href
  {https://doi.org/10.1038/s41467-020-15138-7} {\bibfield  {journal} {\bibinfo
  {journal} {Nature Communications}\ }\textbf {\bibinfo {volume} {11}},\
  \bibinfo {pages} {1605} (\bibinfo {year} {2020})}\BibitemShut {NoStop}%
\bibitem [{\citenamefont {Siyushev}\ \emph {et~al.}(2014)\citenamefont
  {Siyushev}, \citenamefont {Xia}, \citenamefont {Reuter}, \citenamefont
  {Jamali}, \citenamefont {Zhao}, \citenamefont {Yang}, \citenamefont {Duan},
  \citenamefont {Kukharchyk}, \citenamefont {Wieck}, \citenamefont {Kolesov},\
  and\ \citenamefont {Wrachtrup}}]{siyushev14}%
  \BibitemOpen
  \bibfield  {author} {\bibinfo {author} {\bibfnamefont {P.}~\bibnamefont
  {Siyushev}}, \bibinfo {author} {\bibfnamefont {K.}~\bibnamefont {Xia}},
  \bibinfo {author} {\bibfnamefont {R.}~\bibnamefont {Reuter}}, \bibinfo
  {author} {\bibfnamefont {M.}~\bibnamefont {Jamali}}, \bibinfo {author}
  {\bibfnamefont {N.}~\bibnamefont {Zhao}}, \bibinfo {author} {\bibfnamefont
  {N.}~\bibnamefont {Yang}}, \bibinfo {author} {\bibfnamefont {C.}~\bibnamefont
  {Duan}}, \bibinfo {author} {\bibfnamefont {N.}~\bibnamefont {Kukharchyk}},
  \bibinfo {author} {\bibfnamefont {A.~D.}\ \bibnamefont {Wieck}}, \bibinfo
  {author} {\bibfnamefont {R.}~\bibnamefont {Kolesov}},\ and\ \bibinfo {author}
  {\bibfnamefont {J.}~\bibnamefont {Wrachtrup}},\ }\href
  {https://doi.org/10.1038/ncomms4895} {\bibfield  {journal} {\bibinfo
  {journal} {Nature Communications}\ }\textbf {\bibinfo {volume} {5}},\
  \bibinfo {pages} {3895} (\bibinfo {year} {2014})}\BibitemShut {NoStop}%
\bibitem [{\citenamefont {Kindem}\ \emph {et~al.}(2020)\citenamefont {Kindem},
  \citenamefont {Ruskuc}, \citenamefont {Bartholomew}, \citenamefont {Rochman},
  \citenamefont {Huan},\ and\ \citenamefont {Faraon}}]{kindem20}%
  \BibitemOpen
  \bibfield  {author} {\bibinfo {author} {\bibfnamefont {J.~M.}\ \bibnamefont
  {Kindem}}, \bibinfo {author} {\bibfnamefont {A.}~\bibnamefont {Ruskuc}},
  \bibinfo {author} {\bibfnamefont {J.~G.}\ \bibnamefont {Bartholomew}},
  \bibinfo {author} {\bibfnamefont {J.}~\bibnamefont {Rochman}}, \bibinfo
  {author} {\bibfnamefont {Y.~Q.}\ \bibnamefont {Huan}},\ and\ \bibinfo
  {author} {\bibfnamefont {A.}~\bibnamefont {Faraon}},\ }\href
  {https://doi.org/10.1038/s41586-020-2160-9} {\bibfield  {journal} {\bibinfo
  {journal} {Nature}\ }\textbf {\bibinfo {volume} {580}},\ \bibinfo {pages}
  {201} (\bibinfo {year} {2020})}\BibitemShut {NoStop}%
\bibitem [{\citenamefont {Kornher}\ \emph {et~al.}(2020)\citenamefont
  {Kornher}, \citenamefont {Xiao}, \citenamefont {Xia}, \citenamefont {Sardi},
  \citenamefont {Zhao}, \citenamefont {Kolesov},\ and\ \citenamefont
  {Wrachtrup}}]{kornher20}%
  \BibitemOpen
  \bibfield  {author} {\bibinfo {author} {\bibfnamefont {T.}~\bibnamefont
  {Kornher}}, \bibinfo {author} {\bibfnamefont {D.-W.}\ \bibnamefont {Xiao}},
  \bibinfo {author} {\bibfnamefont {K.}~\bibnamefont {Xia}}, \bibinfo {author}
  {\bibfnamefont {F.}~\bibnamefont {Sardi}}, \bibinfo {author} {\bibfnamefont
  {N.}~\bibnamefont {Zhao}}, \bibinfo {author} {\bibfnamefont {R.}~\bibnamefont
  {Kolesov}},\ and\ \bibinfo {author} {\bibfnamefont {J.}~\bibnamefont
  {Wrachtrup}},\ }\href {https://doi.org/10.1103/PhysRevLett.124.170402}
  {\bibfield  {journal} {\bibinfo  {journal} {Physical Review Letters}\
  }\textbf {\bibinfo {volume} {124}},\ \bibinfo {pages} {170402} (\bibinfo
  {year} {2020})}\BibitemShut {NoStop}%
\bibitem [{\citenamefont {B{\"o}ttger}\ \emph {et~al.}(2009)\citenamefont
  {B{\"o}ttger}, \citenamefont {Thiel}, \citenamefont {Cone},\ and\
  \citenamefont {Sun}}]{bottger09}%
  \BibitemOpen
  \bibfield  {author} {\bibinfo {author} {\bibfnamefont {T.}~\bibnamefont
  {B{\"o}ttger}}, \bibinfo {author} {\bibfnamefont {C.~W.}\ \bibnamefont
  {Thiel}}, \bibinfo {author} {\bibfnamefont {R.~L.}\ \bibnamefont {Cone}},\
  and\ \bibinfo {author} {\bibfnamefont {Y.}~\bibnamefont {Sun}},\ }\href
  {https://doi.org/10.1103/PhysRevB.79.115104} {\bibfield  {journal} {\bibinfo
  {journal} {Physical Review B}\ }\textbf {\bibinfo {volume} {79}},\ \bibinfo
  {pages} {115104} (\bibinfo {year} {2009})}\BibitemShut {NoStop}%
\bibitem [{\citenamefont {Ran{\v c}i{\'c}}\ \emph {et~al.}(2018)\citenamefont
  {Ran{\v c}i{\'c}}, \citenamefont {Hedges}, \citenamefont {Ahlefeldt},\ and\
  \citenamefont {Sellars}}]{rancic18}%
  \BibitemOpen
  \bibfield  {author} {\bibinfo {author} {\bibfnamefont {M.}~\bibnamefont
  {Ran{\v c}i{\'c}}}, \bibinfo {author} {\bibfnamefont {M.~P.}\ \bibnamefont
  {Hedges}}, \bibinfo {author} {\bibfnamefont {R.~L.}\ \bibnamefont
  {Ahlefeldt}},\ and\ \bibinfo {author} {\bibfnamefont {M.~J.}\ \bibnamefont
  {Sellars}},\ }\href {https://doi.org/10.1038/nphys4254} {\bibfield  {journal}
  {\bibinfo  {journal} {Nature Physics}\ }\textbf {\bibinfo {volume} {14}},\
  \bibinfo {pages} {50} (\bibinfo {year} {2018})}\BibitemShut {NoStop}%
\bibitem [{\citenamefont {{de Boo}}\ \emph {et~al.}(2020)\citenamefont {{de
  Boo}}, \citenamefont {Yin}, \citenamefont {Ran{\v c}i{\'c}}, \citenamefont
  {Johnson}, \citenamefont {McCallum}, \citenamefont {Sellars},\ and\
  \citenamefont {Rogge}}]{deboo20}%
  \BibitemOpen
  \bibfield  {author} {\bibinfo {author} {\bibfnamefont {G.~G.}\ \bibnamefont
  {{de Boo}}}, \bibinfo {author} {\bibfnamefont {C.}~\bibnamefont {Yin}},
  \bibinfo {author} {\bibfnamefont {M.}~\bibnamefont {Ran{\v c}i{\'c}}},
  \bibinfo {author} {\bibfnamefont {B.~C.}\ \bibnamefont {Johnson}}, \bibinfo
  {author} {\bibfnamefont {J.~C.}\ \bibnamefont {McCallum}}, \bibinfo {author}
  {\bibfnamefont {M.~J.}\ \bibnamefont {Sellars}},\ and\ \bibinfo {author}
  {\bibfnamefont {S.}~\bibnamefont {Rogge}},\ }\href
  {https://doi.org/10.1103/PhysRevB.102.155309} {\bibfield  {journal} {\bibinfo
   {journal} {Physical Review B}\ }\textbf {\bibinfo {volume} {102}},\ \bibinfo
  {pages} {155309} (\bibinfo {year} {2020})}\BibitemShut {NoStop}%
\bibitem [{\citenamefont {Dibos}\ \emph {et~al.}(2018)\citenamefont {Dibos},
  \citenamefont {Raha}, \citenamefont {Phenicie},\ and\ \citenamefont
  {Thompson}}]{dibos18}%
  \BibitemOpen
  \bibfield  {author} {\bibinfo {author} {\bibfnamefont {A.~M.}\ \bibnamefont
  {Dibos}}, \bibinfo {author} {\bibfnamefont {M.}~\bibnamefont {Raha}},
  \bibinfo {author} {\bibfnamefont {C.~M.}\ \bibnamefont {Phenicie}},\ and\
  \bibinfo {author} {\bibfnamefont {J.~D.}\ \bibnamefont {Thompson}},\ }\href
  {https://doi.org/10.1103/PhysRevLett.120.243601} {\bibfield  {journal}
  {\bibinfo  {journal} {Physical Review Letters}\ }\textbf {\bibinfo {volume}
  {120}},\ \bibinfo {pages} {243601} (\bibinfo {year} {2018})}\BibitemShut
  {NoStop}%
\bibitem [{\citenamefont {Carey}\ \emph {et~al.}(1998)\citenamefont {Carey},
  \citenamefont {Barklie}, \citenamefont {Donegan}, \citenamefont {Priolo},
  \citenamefont {Franz{\`o}},\ and\ \citenamefont {Coffa}}]{carey98}%
  \BibitemOpen
  \bibfield  {author} {\bibinfo {author} {\bibfnamefont {J.~D.}\ \bibnamefont
  {Carey}}, \bibinfo {author} {\bibfnamefont {R.~C.}\ \bibnamefont {Barklie}},
  \bibinfo {author} {\bibfnamefont {J.~F.}\ \bibnamefont {Donegan}}, \bibinfo
  {author} {\bibfnamefont {F.}~\bibnamefont {Priolo}}, \bibinfo {author}
  {\bibfnamefont {G.}~\bibnamefont {Franz{\`o}}},\ and\ \bibinfo {author}
  {\bibfnamefont {S.}~\bibnamefont {Coffa}},\ }\href
  {https://doi.org/10.1016/S0022-2313(98)00115-X} {\bibfield  {journal}
  {\bibinfo  {journal} {Journal of Luminescence}\ }\textbf {\bibinfo {volume}
  {80}},\ \bibinfo {pages} {297} (\bibinfo {year} {1998})}\BibitemShut
  {NoStop}%
\bibitem [{\citenamefont {Priolo}\ \emph {et~al.}(1995)\citenamefont {Priolo},
  \citenamefont {Franz{\`o}}, \citenamefont {Coffa}, \citenamefont {Polman},
  \citenamefont {Libertino}, \citenamefont {Barklie},\ and\ \citenamefont
  {Carey}}]{priolo95}%
  \BibitemOpen
  \bibfield  {author} {\bibinfo {author} {\bibfnamefont {F.}~\bibnamefont
  {Priolo}}, \bibinfo {author} {\bibfnamefont {G.}~\bibnamefont {Franz{\`o}}},
  \bibinfo {author} {\bibfnamefont {S.}~\bibnamefont {Coffa}}, \bibinfo
  {author} {\bibfnamefont {A.}~\bibnamefont {Polman}}, \bibinfo {author}
  {\bibfnamefont {S.}~\bibnamefont {Libertino}}, \bibinfo {author}
  {\bibfnamefont {R.}~\bibnamefont {Barklie}},\ and\ \bibinfo {author}
  {\bibfnamefont {D.}~\bibnamefont {Carey}},\ }\href
  {https://doi.org/10.1063/1.359904} {\bibfield  {journal} {\bibinfo  {journal}
  {Journal of Applied Physics}\ }\textbf {\bibinfo {volume} {78}},\ \bibinfo
  {pages} {3874} (\bibinfo {year} {1995})}\BibitemShut {NoStop}%
\bibitem [{\citenamefont {Liu}\ \emph {et~al.}(1995)\citenamefont {Liu},
  \citenamefont {Zhang}, \citenamefont {Wilson}, \citenamefont {Curello},
  \citenamefont {Rao},\ and\ \citenamefont {Hemment}}]{liu95}%
  \BibitemOpen
  \bibfield  {author} {\bibinfo {author} {\bibfnamefont {P.}~\bibnamefont
  {Liu}}, \bibinfo {author} {\bibfnamefont {J.~P.}\ \bibnamefont {Zhang}},
  \bibinfo {author} {\bibfnamefont {R.~J.}\ \bibnamefont {Wilson}}, \bibinfo
  {author} {\bibfnamefont {G.}~\bibnamefont {Curello}}, \bibinfo {author}
  {\bibfnamefont {S.~S.}\ \bibnamefont {Rao}},\ and\ \bibinfo {author}
  {\bibfnamefont {P.~L.~F.}\ \bibnamefont {Hemment}},\ }\href
  {https://doi.org/10.1063/1.113708} {\bibfield  {journal} {\bibinfo  {journal}
  {Applied Physics Letters}\ }\textbf {\bibinfo {volume} {66}},\ \bibinfo
  {pages} {3158} (\bibinfo {year} {1995})}\BibitemShut {NoStop}%
\bibitem [{\citenamefont {Hughes}\ \emph {et~al.}(2021)\citenamefont {Hughes},
  \citenamefont {Panjwani}, \citenamefont {Urdampilleta}, \citenamefont
  {Homewood}, \citenamefont {Murdin},\ and\ \citenamefont {Carey}}]{hughes21}%
  \BibitemOpen
  \bibfield  {author} {\bibinfo {author} {\bibfnamefont {M.~A.}\ \bibnamefont
  {Hughes}}, \bibinfo {author} {\bibfnamefont {N.~A.}\ \bibnamefont
  {Panjwani}}, \bibinfo {author} {\bibfnamefont {M.}~\bibnamefont
  {Urdampilleta}}, \bibinfo {author} {\bibfnamefont {K.~P.}\ \bibnamefont
  {Homewood}}, \bibinfo {author} {\bibfnamefont {B.}~\bibnamefont {Murdin}},\
  and\ \bibinfo {author} {\bibfnamefont {J.~D.}\ \bibnamefont {Carey}},\ }\href
  {https://doi.org/10.1063/5.0046904} {\bibfield  {journal} {\bibinfo
  {journal} {Applied Physics Letters}\ }\textbf {\bibinfo {volume} {118}},\
  \bibinfo {pages} {194001} (\bibinfo {year} {2021})}\BibitemShut {NoStop}%
\bibitem [{\citenamefont {Przybylinska}\ \emph {et~al.}(1996)\citenamefont
  {Przybylinska}, \citenamefont {Jantsch}, \citenamefont {{Suprun-Belevitch}},
  \citenamefont {Stepikhova}, \citenamefont {Palmetshofer}, \citenamefont
  {Hendorfer}, \citenamefont {Kozanecki}, \citenamefont {Wilson},\ and\
  \citenamefont {Sealy}}]{przybylinska96}%
  \BibitemOpen
  \bibfield  {author} {\bibinfo {author} {\bibfnamefont {H.}~\bibnamefont
  {Przybylinska}}, \bibinfo {author} {\bibfnamefont {W.}~\bibnamefont
  {Jantsch}}, \bibinfo {author} {\bibfnamefont {Y.}~\bibnamefont
  {{Suprun-Belevitch}}}, \bibinfo {author} {\bibfnamefont {M.}~\bibnamefont
  {Stepikhova}}, \bibinfo {author} {\bibfnamefont {L.}~\bibnamefont
  {Palmetshofer}}, \bibinfo {author} {\bibfnamefont {G.}~\bibnamefont
  {Hendorfer}}, \bibinfo {author} {\bibfnamefont {A.}~\bibnamefont
  {Kozanecki}}, \bibinfo {author} {\bibfnamefont {R.~J.}\ \bibnamefont
  {Wilson}},\ and\ \bibinfo {author} {\bibfnamefont {B.~J.}\ \bibnamefont
  {Sealy}},\ }\href {https://doi.org/10.1103/PhysRevB.54.2532} {\bibfield
  {journal} {\bibinfo  {journal} {Physical Review B}\ }\textbf {\bibinfo
  {volume} {54}},\ \bibinfo {pages} {2532} (\bibinfo {year}
  {1996})}\BibitemShut {NoStop}%
\bibitem [{\citenamefont {Tang}\ \emph {et~al.}(1989)\citenamefont {Tang},
  \citenamefont {Heasman}, \citenamefont {Gillin},\ and\ \citenamefont
  {Sealy}}]{tang89}%
  \BibitemOpen
  \bibfield  {author} {\bibinfo {author} {\bibfnamefont {Y.~S.}\ \bibnamefont
  {Tang}}, \bibinfo {author} {\bibfnamefont {K.~C.}\ \bibnamefont {Heasman}},
  \bibinfo {author} {\bibfnamefont {W.~P.}\ \bibnamefont {Gillin}},\ and\
  \bibinfo {author} {\bibfnamefont {B.~J.}\ \bibnamefont {Sealy}},\ }\href
  {https://doi.org/10.1063/1.101888} {\bibfield  {journal} {\bibinfo  {journal}
  {Applied Physics Letters}\ }\textbf {\bibinfo {volume} {55}},\ \bibinfo
  {pages} {432} (\bibinfo {year} {1989})}\BibitemShut {NoStop}%
\bibitem [{\citenamefont {Vinh}\ \emph {et~al.}(2003)\citenamefont {Vinh},
  \citenamefont {Przybyli{\'n}ska}, \citenamefont {Krasil'nik},\ and\
  \citenamefont {Gregorkiewicz}}]{vinh03}%
  \BibitemOpen
  \bibfield  {author} {\bibinfo {author} {\bibfnamefont {N.~Q.}\ \bibnamefont
  {Vinh}}, \bibinfo {author} {\bibfnamefont {H.}~\bibnamefont
  {Przybyli{\'n}ska}}, \bibinfo {author} {\bibfnamefont {Z.~F.}\ \bibnamefont
  {Krasil'nik}},\ and\ \bibinfo {author} {\bibfnamefont {T.}~\bibnamefont
  {Gregorkiewicz}},\ }\href {https://doi.org/10.1103/PhysRevLett.90.066401}
  {\bibfield  {journal} {\bibinfo  {journal} {Physical Review Letters}\
  }\textbf {\bibinfo {volume} {90}},\ \bibinfo {pages} {066401} (\bibinfo
  {year} {2003})}\BibitemShut {NoStop}%
\bibitem [{\citenamefont {Weiss}\ \emph {et~al.}(2021)\citenamefont {Weiss},
  \citenamefont {Gritsch}, \citenamefont {Merkel},\ and\ \citenamefont
  {Reiserer}}]{weiss21}%
  \BibitemOpen
  \bibfield  {author} {\bibinfo {author} {\bibfnamefont {L.}~\bibnamefont
  {Weiss}}, \bibinfo {author} {\bibfnamefont {A.}~\bibnamefont {Gritsch}},
  \bibinfo {author} {\bibfnamefont {B.}~\bibnamefont {Merkel}},\ and\ \bibinfo
  {author} {\bibfnamefont {A.}~\bibnamefont {Reiserer}},\ }\href
  {https://doi.org/10.1364/OPTICA.413330} {\bibfield  {journal} {\bibinfo
  {journal} {Optica}\ }\textbf {\bibinfo {volume} {8}},\ \bibinfo {pages} {40}
  (\bibinfo {year} {2021})}\BibitemShut {NoStop}%
\bibitem [{\citenamefont {Berkman}\ \emph {et~al.}(tion)\citenamefont
  {Berkman}, \citenamefont {{de Boo}}, \citenamefont {Lyasota}, \citenamefont
  {Bartholomew}, \citenamefont {Johnson}, \citenamefont {McCallum},
  \citenamefont {Xu}, \citenamefont {Xie}, \citenamefont {Ahlefeldt},
  \citenamefont {Sellars}, \citenamefont {Yin},\ and\ \citenamefont
  {Rogge}}]{berkman21}%
  \BibitemOpen
  \bibfield  {author} {\bibinfo {author} {\bibfnamefont {I.}~\bibnamefont
  {Berkman}}, \bibinfo {author} {\bibfnamefont {G.}~\bibnamefont {{de Boo}}},
  \bibinfo {author} {\bibfnamefont {A.}~\bibnamefont {Lyasota}}, \bibinfo
  {author} {\bibfnamefont {J.~G.}\ \bibnamefont {Bartholomew}}, \bibinfo
  {author} {\bibfnamefont {B.~C.}\ \bibnamefont {Johnson}}, \bibinfo {author}
  {\bibfnamefont {J.}~\bibnamefont {McCallum}}, \bibinfo {author}
  {\bibfnamefont {B.-B.}\ \bibnamefont {Xu}}, \bibinfo {author} {\bibfnamefont
  {S.}~\bibnamefont {Xie}}, \bibinfo {author} {\bibfnamefont {R.}~\bibnamefont
  {Ahlefeldt}}, \bibinfo {author} {\bibfnamefont {M.}~\bibnamefont {Sellars}},
  \bibinfo {author} {\bibfnamefont {C.}~\bibnamefont {Yin}},\ and\ \bibinfo
  {author} {\bibfnamefont {S.}~\bibnamefont {Rogge}}} (\bibinfo {year} {in
  preparation})\BibitemShut {NoStop}%
\bibitem [{\citenamefont {Carey}\ \emph {et~al.}(1996)\citenamefont {Carey},
  \citenamefont {Donegan}, \citenamefont {Barklie}, \citenamefont {Priolo},
  \citenamefont {Franz{\`o}},\ and\ \citenamefont {Coffa}}]{carey96}%
  \BibitemOpen
  \bibfield  {author} {\bibinfo {author} {\bibfnamefont {J.~D.}\ \bibnamefont
  {Carey}}, \bibinfo {author} {\bibfnamefont {J.~F.}\ \bibnamefont {Donegan}},
  \bibinfo {author} {\bibfnamefont {R.~C.}\ \bibnamefont {Barklie}}, \bibinfo
  {author} {\bibfnamefont {F.}~\bibnamefont {Priolo}}, \bibinfo {author}
  {\bibfnamefont {G.}~\bibnamefont {Franz{\`o}}},\ and\ \bibinfo {author}
  {\bibfnamefont {S.}~\bibnamefont {Coffa}},\ }\href
  {https://doi.org/10.1063/1.117127} {\bibfield  {journal} {\bibinfo  {journal}
  {Applied Physics Letters}\ }\textbf {\bibinfo {volume} {69}},\ \bibinfo
  {pages} {3854} (\bibinfo {year} {1996})}\BibitemShut {NoStop}%
\bibitem [{\citenamefont {Carey}\ \emph {et~al.}(1999)\citenamefont {Carey},
  \citenamefont {Barklie}, \citenamefont {Donegan}, \citenamefont {Priolo},
  \citenamefont {Franz{\`o}},\ and\ \citenamefont {Coffa}}]{carey99}%
  \BibitemOpen
  \bibfield  {author} {\bibinfo {author} {\bibfnamefont {J.~D.}\ \bibnamefont
  {Carey}}, \bibinfo {author} {\bibfnamefont {R.~C.}\ \bibnamefont {Barklie}},
  \bibinfo {author} {\bibfnamefont {J.~F.}\ \bibnamefont {Donegan}}, \bibinfo
  {author} {\bibfnamefont {F.}~\bibnamefont {Priolo}}, \bibinfo {author}
  {\bibfnamefont {G.}~\bibnamefont {Franz{\`o}}},\ and\ \bibinfo {author}
  {\bibfnamefont {S.}~\bibnamefont {Coffa}},\ }\href
  {https://doi.org/10.1103/PhysRevB.59.2773} {\bibfield  {journal} {\bibinfo
  {journal} {Physical Review B}\ }\textbf {\bibinfo {volume} {59}},\ \bibinfo
  {pages} {2773} (\bibinfo {year} {1999})}\BibitemShut {NoStop}%
\bibitem [{\citenamefont {Carey}(2002)}]{carey02}%
  \BibitemOpen
  \bibfield  {author} {\bibinfo {author} {\bibfnamefont {J.~D.}\ \bibnamefont
  {Carey}},\ }\href {https://doi.org/10.1088/0953-8984/14/36/310} {\bibfield
  {journal} {\bibinfo  {journal} {Journal of Physics: Condensed Matter}\
  }\textbf {\bibinfo {volume} {14}},\ \bibinfo {pages} {8537} (\bibinfo {year}
  {2002})}\BibitemShut {NoStop}%
\bibitem [{\citenamefont {Hughes}\ \emph {et~al.}(2019)\citenamefont {Hughes},
  \citenamefont {Li}, \citenamefont {Theodoropoulou},\ and\ \citenamefont
  {Carey}}]{hughes19}%
  \BibitemOpen
  \bibfield  {author} {\bibinfo {author} {\bibfnamefont {M.~A.}\ \bibnamefont
  {Hughes}}, \bibinfo {author} {\bibfnamefont {H.}~\bibnamefont {Li}}, \bibinfo
  {author} {\bibfnamefont {N.}~\bibnamefont {Theodoropoulou}},\ and\ \bibinfo
  {author} {\bibfnamefont {J.~D.}\ \bibnamefont {Carey}},\ }\href
  {https://doi.org/10.1038/s41598-019-55246-z} {\bibfield  {journal} {\bibinfo
  {journal} {Scientific Reports}\ }\textbf {\bibinfo {volume} {9}},\ \bibinfo
  {pages} {19031} (\bibinfo {year} {2019})}\BibitemShut {NoStop}%
\bibitem [{\citenamefont {{Guillot-No{\"e}l}}\ \emph
  {et~al.}(2004)\citenamefont {{Guillot-No{\"e}l}}, \citenamefont {Goldner},
  \citenamefont {Higel},\ and\ \citenamefont {Gourier}}]{guillot-noel04}%
  \BibitemOpen
  \bibfield  {author} {\bibinfo {author} {\bibfnamefont {O.}~\bibnamefont
  {{Guillot-No{\"e}l}}}, \bibinfo {author} {\bibfnamefont {P.}~\bibnamefont
  {Goldner}}, \bibinfo {author} {\bibfnamefont {P.}~\bibnamefont {Higel}},\
  and\ \bibinfo {author} {\bibfnamefont {D.}~\bibnamefont {Gourier}},\
  }\href@noop {} {\bibfield  {journal} {\bibinfo  {journal} {Journal of
  Physics: Condensed Matter}\ }\textbf {\bibinfo {volume} {16}},\ \bibinfo
  {pages} {R1} (\bibinfo {year} {2004})}\BibitemShut {NoStop}%
\bibitem [{\citenamefont {Cone}\ and\ \citenamefont {Meltzer}(1987)}]{cone87}%
  \BibitemOpen
  \bibfield  {author} {\bibinfo {author} {\bibfnamefont {R.~L.}\ \bibnamefont
  {Cone}}\ and\ \bibinfo {author} {\bibfnamefont {R.~S.}\ \bibnamefont
  {Meltzer}},\ }in\ \href@noop {} {\emph {\bibinfo {booktitle} {Spectroscopy of
  {{Solids Containing Rare Earth Ions}}}}},\ \bibinfo {editor} {edited by\
  \bibinfo {editor} {\bibfnamefont {A.~A.}\ \bibnamefont {Kaplyanskii}}\ and\
  \bibinfo {editor} {\bibfnamefont {R.~M.}\ \bibnamefont {Macfarlane}}}\
  (\bibinfo  {publisher} {North Holland},\ \bibinfo {year} {1987})\ p.\
  \bibinfo {pages} {481}\BibitemShut {NoStop}%
\bibitem [{\citenamefont {Fricke}(1979)}]{fricke79a}%
  \BibitemOpen
  \bibfield  {author} {\bibinfo {author} {\bibfnamefont {W.}~\bibnamefont
  {Fricke}},\ }\href {https://doi.org/10.1007/BF01323502} {\bibfield  {journal}
  {\bibinfo  {journal} {Zeitschrift f\"ur Physik B Condensed Matter}\ }\textbf
  {\bibinfo {volume} {33}},\ \bibinfo {pages} {261} (\bibinfo {year}
  {1979})}\BibitemShut {NoStop}%
\bibitem [{\citenamefont {Yamaguchi}\ \emph {et~al.}(1998)\citenamefont
  {Yamaguchi}, \citenamefont {Koyama}, \citenamefont {Suemoto},\ and\
  \citenamefont {Mitsunaga}}]{yamaguchi98}%
  \BibitemOpen
  \bibfield  {author} {\bibinfo {author} {\bibfnamefont {M.}~\bibnamefont
  {Yamaguchi}}, \bibinfo {author} {\bibfnamefont {K.}~\bibnamefont {Koyama}},
  \bibinfo {author} {\bibfnamefont {T.}~\bibnamefont {Suemoto}},\ and\ \bibinfo
  {author} {\bibfnamefont {M.}~\bibnamefont {Mitsunaga}},\ }\href
  {https://doi.org/10.1016/S0022-2313(97)00267-6} {\bibfield  {journal}
  {\bibinfo  {journal} {Journal of Luminescence}\ }\textbf {\bibinfo {volume}
  {76-77}},\ \bibinfo {pages} {681} (\bibinfo {year} {1998})}\BibitemShut
  {NoStop}%
\bibitem [{\citenamefont {Ahlefeldt}\ \emph
  {et~al.}(2013{\natexlab{a}})\citenamefont {Ahlefeldt}, \citenamefont
  {Hutchison}, \citenamefont {Manson},\ and\ \citenamefont
  {Sellars}}]{ahlefeldt13method}%
  \BibitemOpen
  \bibfield  {author} {\bibinfo {author} {\bibfnamefont {R.~L.}\ \bibnamefont
  {Ahlefeldt}}, \bibinfo {author} {\bibfnamefont {W.~D.}\ \bibnamefont
  {Hutchison}}, \bibinfo {author} {\bibfnamefont {N.~B.}\ \bibnamefont
  {Manson}},\ and\ \bibinfo {author} {\bibfnamefont {M.~J.}\ \bibnamefont
  {Sellars}},\ }\href {https://doi.org/10.1103/PhysRevB.88.184424} {\bibfield
  {journal} {\bibinfo  {journal} {Physical Review B}\ }\textbf {\bibinfo
  {volume} {88}},\ \bibinfo {pages} {184424} (\bibinfo {year}
  {2013}{\natexlab{a}})}\BibitemShut {NoStop}%
\bibitem [{\citenamefont {Laplane}\ \emph {et~al.}(2016)\citenamefont
  {Laplane}, \citenamefont {Zambrini~Cruzeiro}, \citenamefont {Fr{\"o}wis},
  \citenamefont {Goldner},\ and\ \citenamefont {Afzelius}}]{laplane16high}%
  \BibitemOpen
  \bibfield  {author} {\bibinfo {author} {\bibfnamefont {C.}~\bibnamefont
  {Laplane}}, \bibinfo {author} {\bibfnamefont {E.}~\bibnamefont
  {Zambrini~Cruzeiro}}, \bibinfo {author} {\bibfnamefont {F.}~\bibnamefont
  {Fr{\"o}wis}}, \bibinfo {author} {\bibfnamefont {P.}~\bibnamefont
  {Goldner}},\ and\ \bibinfo {author} {\bibfnamefont {M.}~\bibnamefont
  {Afzelius}},\ }\href {https://doi.org/10.1103/PhysRevLett.117.037203}
  {\bibfield  {journal} {\bibinfo  {journal} {Physical Review Letters}\
  }\textbf {\bibinfo {volume} {117}},\ \bibinfo {pages} {037203} (\bibinfo
  {year} {2016})}\BibitemShut {NoStop}%
\bibitem [{\citenamefont {Wolf}(1971)}]{wolf71}%
  \BibitemOpen
  \bibfield  {author} {\bibinfo {author} {\bibfnamefont {W.~P.}\ \bibnamefont
  {Wolf}},\ }\href {https://doi.org/10.1051/jphyscol:1971106} {\bibfield
  {journal} {\bibinfo  {journal} {Le Journal de Physique Colloques}\ }\textbf
  {\bibinfo {volume} {32}},\ \bibinfo {pages} {C1} (\bibinfo {year}
  {1971})}\BibitemShut {NoStop}%
\bibitem [{\citenamefont {Baker}(1971)}]{baker71}%
  \BibitemOpen
  \bibfield  {author} {\bibinfo {author} {\bibfnamefont {J.~M.}\ \bibnamefont
  {Baker}},\ }\href {https://doi.org/10.1088/0034-4885/34/1/303} {\bibfield
  {journal} {\bibinfo  {journal} {Reports on Progress in Physics}\ }\textbf
  {\bibinfo {volume} {34}},\ \bibinfo {pages} {109} (\bibinfo {year}
  {1971})}\BibitemShut {NoStop}%
\bibitem [{\citenamefont {Ohlsson}\ \emph {et~al.}(2002)\citenamefont
  {Ohlsson}, \citenamefont {Krishna~Mohan},\ and\ \citenamefont
  {Kr{\"o}ll}}]{ohlsson02}%
  \BibitemOpen
  \bibfield  {author} {\bibinfo {author} {\bibfnamefont {N.}~\bibnamefont
  {Ohlsson}}, \bibinfo {author} {\bibfnamefont {R.}~\bibnamefont
  {Krishna~Mohan}},\ and\ \bibinfo {author} {\bibfnamefont {S.}~\bibnamefont
  {Kr{\"o}ll}},\ }\href {https://doi.org/10.1016/S0030-4018(01)01666-2}
  {\bibfield  {journal} {\bibinfo  {journal} {Optics Communications}\ }\textbf
  {\bibinfo {volume} {201}},\ \bibinfo {pages} {71} (\bibinfo {year}
  {2002})}\BibitemShut {NoStop}%
\bibitem [{\citenamefont {Grimm}\ \emph {et~al.}(2021)\citenamefont {Grimm},
  \citenamefont {Beckert}, \citenamefont {Aeppli},\ and\ \citenamefont
  {M{\"u}ller}}]{grimm21}%
  \BibitemOpen
  \bibfield  {author} {\bibinfo {author} {\bibfnamefont {M.}~\bibnamefont
  {Grimm}}, \bibinfo {author} {\bibfnamefont {A.}~\bibnamefont {Beckert}},
  \bibinfo {author} {\bibfnamefont {G.}~\bibnamefont {Aeppli}},\ and\ \bibinfo
  {author} {\bibfnamefont {M.}~\bibnamefont {M{\"u}ller}},\ }\href
  {https://doi.org/10.1103/PRXQuantum.2.010312} {\bibfield  {journal} {\bibinfo
   {journal} {PRX Quantum}\ }\textbf {\bibinfo {volume} {2}},\ \bibinfo {pages}
  {010312} (\bibinfo {year} {2021})}\BibitemShut {NoStop}%
\bibitem [{\citenamefont {Ahlefeldt}\ \emph {et~al.}(2020)\citenamefont
  {Ahlefeldt}, \citenamefont {Pearce}, \citenamefont {Hush},\ and\
  \citenamefont {Sellars}}]{ahlefeldt20}%
  \BibitemOpen
  \bibfield  {author} {\bibinfo {author} {\bibfnamefont {R.~L.}\ \bibnamefont
  {Ahlefeldt}}, \bibinfo {author} {\bibfnamefont {M.~J.}\ \bibnamefont
  {Pearce}}, \bibinfo {author} {\bibfnamefont {M.~R.}\ \bibnamefont {Hush}},\
  and\ \bibinfo {author} {\bibfnamefont {M.~J.}\ \bibnamefont {Sellars}},\
  }\href {https://doi.org/10.1103/PhysRevA.101.012309} {\bibfield  {journal}
  {\bibinfo  {journal} {Physical Review A}\ }\textbf {\bibinfo {volume}
  {101}},\ \bibinfo {pages} {012309} (\bibinfo {year} {2020})}\BibitemShut
  {NoStop}%
\bibitem [{\citenamefont {Ahlefeldt}\ \emph
  {et~al.}(2013{\natexlab{b}})\citenamefont {Ahlefeldt}, \citenamefont
  {McAuslan}, \citenamefont {Longdell}, \citenamefont {Manson},\ and\
  \citenamefont {Sellars}}]{ahlefeldt13precision}%
  \BibitemOpen
  \bibfield  {author} {\bibinfo {author} {\bibfnamefont {R.~L.}\ \bibnamefont
  {Ahlefeldt}}, \bibinfo {author} {\bibfnamefont {D.~L.}\ \bibnamefont
  {McAuslan}}, \bibinfo {author} {\bibfnamefont {J.~J.}\ \bibnamefont
  {Longdell}}, \bibinfo {author} {\bibfnamefont {N.~B.}\ \bibnamefont
  {Manson}},\ and\ \bibinfo {author} {\bibfnamefont {M.~J.}\ \bibnamefont
  {Sellars}},\ }\href {https://doi.org/10.1103/PhysRevLett.111.240501}
  {\bibfield  {journal} {\bibinfo  {journal} {Physical Review Letters}\
  }\textbf {\bibinfo {volume} {111}},\ \bibinfo {pages} {240501} (\bibinfo
  {year} {2013}{\natexlab{b}})}\BibitemShut {NoStop}%
\bibitem [{\citenamefont {Lukac}\ and\ \citenamefont {Hahn}(1989)}]{lukac89}%
  \BibitemOpen
  \bibfield  {author} {\bibinfo {author} {\bibfnamefont {M.}~\bibnamefont
  {Lukac}}\ and\ \bibinfo {author} {\bibfnamefont {E.}~\bibnamefont {Hahn}},\
  }\href {https://doi.org/10.1016/0030-4018(89)90064-3} {\bibfield  {journal}
  {\bibinfo  {journal} {Optics Communications}\ }\textbf {\bibinfo {volume}
  {70}},\ \bibinfo {pages} {195} (\bibinfo {year} {1989})}\BibitemShut
  {NoStop}%
\bibitem [{\citenamefont {{Guillot-No{\"e}l}}\ \emph
  {et~al.}(2000)\citenamefont {{Guillot-No{\"e}l}}, \citenamefont {Mehta},
  \citenamefont {Viana}, \citenamefont {Gourier}, \citenamefont {Boukhris},\
  and\ \citenamefont {Jandl}}]{guillot-noel00}%
  \BibitemOpen
  \bibfield  {author} {\bibinfo {author} {\bibfnamefont {O.}~\bibnamefont
  {{Guillot-No{\"e}l}}}, \bibinfo {author} {\bibfnamefont {V.}~\bibnamefont
  {Mehta}}, \bibinfo {author} {\bibfnamefont {B.}~\bibnamefont {Viana}},
  \bibinfo {author} {\bibfnamefont {D.}~\bibnamefont {Gourier}}, \bibinfo
  {author} {\bibfnamefont {M.}~\bibnamefont {Boukhris}},\ and\ \bibinfo
  {author} {\bibfnamefont {S.}~\bibnamefont {Jandl}},\ }\href
  {https://doi.org/10.1103/PhysRevB.61.15338} {\bibfield  {journal} {\bibinfo
  {journal} {Physical Review B}\ }\textbf {\bibinfo {volume} {61}},\ \bibinfo
  {pages} {15338} (\bibinfo {year} {2000})}\BibitemShut {NoStop}%
\bibitem [{\citenamefont {Clemens}\ and\ \citenamefont
  {Hutchison}(1983)}]{clemens83}%
  \BibitemOpen
  \bibfield  {author} {\bibinfo {author} {\bibfnamefont {J.~M.}\ \bibnamefont
  {Clemens}}\ and\ \bibinfo {author} {\bibfnamefont {C.~A.}\ \bibnamefont
  {Hutchison}},\ }\href {https://doi.org/10.1103/PhysRevB.28.50} {\bibfield
  {journal} {\bibinfo  {journal} {Physical Review B}\ }\textbf {\bibinfo
  {volume} {28}},\ \bibinfo {pages} {50} (\bibinfo {year} {1983})}\BibitemShut
  {NoStop}%
\bibitem [{\citenamefont {Erd{\"o}s}(1966)}]{erdos66}%
  \BibitemOpen
  \bibfield  {author} {\bibinfo {author} {\bibfnamefont {P.}~\bibnamefont
  {Erd{\"o}s}},\ }\href {https://doi.org/10.1016/0022-3697(66)90100-4}
  {\bibfield  {journal} {\bibinfo  {journal} {Journal of Physics and Chemistry
  of Solids}\ }\textbf {\bibinfo {volume} {27}},\ \bibinfo {pages} {1705}
  (\bibinfo {year} {1966})}\BibitemShut {NoStop}%
\bibitem [{\citenamefont {Carey}(2009)}]{carey09}%
  \BibitemOpen
  \bibfield  {author} {\bibinfo {author} {\bibfnamefont {J.~D.}\ \bibnamefont
  {Carey}},\ }\href {https://doi.org/10.1088/0953-8984/21/17/175601} {\bibfield
   {journal} {\bibinfo  {journal} {Journal of Physics: Condensed Matter}\
  }\textbf {\bibinfo {volume} {21}},\ \bibinfo {pages} {175601} (\bibinfo
  {year} {2009})}\BibitemShut {NoStop}%
\bibitem [{\citenamefont {Wolf}(2000)}]{wolf00}%
  \BibitemOpen
  \bibfield  {author} {\bibinfo {author} {\bibfnamefont {W.~P.}\ \bibnamefont
  {Wolf}},\ }\href {https://doi.org/10.1590/S0103-97332000000400030} {\bibfield
   {journal} {\bibinfo  {journal} {Brazilian Journal of Physics}\ }\textbf
  {\bibinfo {volume} {30}},\ \bibinfo {pages} {794} (\bibinfo {year}
  {2000})}\BibitemShut {NoStop}%
\bibitem [{\citenamefont {Kenyon}(2005)}]{kenyon05}%
  \BibitemOpen
  \bibfield  {author} {\bibinfo {author} {\bibfnamefont {A.~J.}\ \bibnamefont
  {Kenyon}},\ }\href {https://doi.org/10.1088/0268-1242/20/12/R02} {\bibfield
  {journal} {\bibinfo  {journal} {Semiconductor Science and Technology}\
  }\textbf {\bibinfo {volume} {20}},\ \bibinfo {pages} {R65} (\bibinfo {year}
  {2005})}\BibitemShut {NoStop}%
\bibitem [{\citenamefont {Birgeneau}(1968)}]{birgeneau68}%
  \BibitemOpen
  \bibfield  {author} {\bibinfo {author} {\bibfnamefont {R.~J.}\ \bibnamefont
  {Birgeneau}},\ }\href {https://doi.org/doi:10.1063/1.1652567} {\bibfield
  {journal} {\bibinfo  {journal} {Applied Physics Letters}\ }\textbf {\bibinfo
  {volume} {13}},\ \bibinfo {pages} {193} (\bibinfo {year} {1968})}\BibitemShut
  {NoStop}%
\bibitem [{\citenamefont {Brower}\ \emph {et~al.}(1966)\citenamefont {Brower},
  \citenamefont {Stapleton},\ and\ \citenamefont {Brower}}]{brower66}%
  \BibitemOpen
  \bibfield  {author} {\bibinfo {author} {\bibfnamefont {K.~L.}\ \bibnamefont
  {Brower}}, \bibinfo {author} {\bibfnamefont {H.~J.}\ \bibnamefont
  {Stapleton}},\ and\ \bibinfo {author} {\bibfnamefont {E.~O.}\ \bibnamefont
  {Brower}},\ }\href {https://doi.org/10.1103/PhysRev.146.233} {\bibfield
  {journal} {\bibinfo  {journal} {Physical Review}\ }\textbf {\bibinfo {volume}
  {146}},\ \bibinfo {pages} {233} (\bibinfo {year} {1966})}\BibitemShut
  {NoStop}%
\bibitem [{\citenamefont {Cone}\ and\ \citenamefont {Meltzer}(1973)}]{cone73}%
  \BibitemOpen
  \bibfield  {author} {\bibinfo {author} {\bibfnamefont {R.~L.}\ \bibnamefont
  {Cone}}\ and\ \bibinfo {author} {\bibfnamefont {R.~S.}\ \bibnamefont
  {Meltzer}},\ }\href {https://doi.org/10.1103/PhysRevLett.30.859} {\bibfield
  {journal} {\bibinfo  {journal} {Physical Review Letters}\ }\textbf {\bibinfo
  {volume} {30}},\ \bibinfo {pages} {859} (\bibinfo {year} {1973})}\BibitemShut
  {NoStop}%
\bibitem [{\citenamefont {Cone}\ and\ \citenamefont {Meltzer}(1975)}]{cone75}%
  \BibitemOpen
  \bibfield  {author} {\bibinfo {author} {\bibfnamefont {R.~L.}\ \bibnamefont
  {Cone}}\ and\ \bibinfo {author} {\bibfnamefont {R.~S.}\ \bibnamefont
  {Meltzer}},\ }\href {https://doi.org/10.1063/1.430951} {\bibfield  {journal}
  {\bibinfo  {journal} {The Journal of Chemical Physics}\ }\textbf {\bibinfo
  {volume} {62}},\ \bibinfo {pages} {3573} (\bibinfo {year}
  {1975})}\BibitemShut {NoStop}%
\bibitem [{\citenamefont {Prinz}(1966)}]{prinz66}%
  \BibitemOpen
  \bibfield  {author} {\bibinfo {author} {\bibfnamefont {G.~A.}\ \bibnamefont
  {Prinz}},\ }\href {https://doi.org/10.1103/PhysRev.152.474} {\bibfield
  {journal} {\bibinfo  {journal} {Physical Review}\ }\textbf {\bibinfo {volume}
  {152}},\ \bibinfo {pages} {474} (\bibinfo {year} {1966})}\BibitemShut
  {NoStop}%
\bibitem [{\citenamefont {Abraham}\ \emph {et~al.}(1992)\citenamefont
  {Abraham}, \citenamefont {Baker}, \citenamefont {Bleaney}, \citenamefont
  {Pfeffer},\ and\ \citenamefont {Wells}}]{abraham92}%
  \BibitemOpen
  \bibfield  {author} {\bibinfo {author} {\bibfnamefont {M.~M.}\ \bibnamefont
  {Abraham}}, \bibinfo {author} {\bibfnamefont {J.~M.}\ \bibnamefont {Baker}},
  \bibinfo {author} {\bibfnamefont {B.}~\bibnamefont {Bleaney}}, \bibinfo
  {author} {\bibfnamefont {J.~Z.}\ \bibnamefont {Pfeffer}},\ and\ \bibinfo
  {author} {\bibfnamefont {M.~R.}\ \bibnamefont {Wells}},\ }\href
  {https://doi.org/10.1088/0953-8984/4/24/014} {\bibfield  {journal} {\bibinfo
  {journal} {Journal of Physics: Condensed Matter}\ }\textbf {\bibinfo {volume}
  {4}},\ \bibinfo {pages} {5443} (\bibinfo {year} {1992})}\BibitemShut
  {NoStop}%
\bibitem [{\citenamefont {Longdell}\ \emph {et~al.}(2004)\citenamefont
  {Longdell}, \citenamefont {Sellars},\ and\ \citenamefont
  {Manson}}]{longdell04a}%
  \BibitemOpen
  \bibfield  {author} {\bibinfo {author} {\bibfnamefont {J.~J.}\ \bibnamefont
  {Longdell}}, \bibinfo {author} {\bibfnamefont {M.~J.}\ \bibnamefont
  {Sellars}},\ and\ \bibinfo {author} {\bibfnamefont {N.~B.}\ \bibnamefont
  {Manson}},\ }\href {https://doi.org/10.1103/PhysRevLett.93.130503} {\bibfield
   {journal} {\bibinfo  {journal} {Physical Review Letters}\ }\textbf {\bibinfo
  {volume} {93}},\ \bibinfo {pages} {130503} (\bibinfo {year}
  {2004})}\BibitemShut {NoStop}%
\bibitem [{\citenamefont {Kinos}\ \emph {et~al.}(2021)\citenamefont {Kinos},
  \citenamefont {Hunger}, \citenamefont {Kolesov}, \citenamefont {M{\o}lmer},
  \citenamefont {{de Riedmatten}}, \citenamefont {Goldner}, \citenamefont
  {Tallaire}, \citenamefont {Morvan}, \citenamefont {Berger}, \citenamefont
  {Welinski}, \citenamefont {Karrai}, \citenamefont {Rippe}, \citenamefont
  {Kr{\"o}ll},\ and\ \citenamefont {Walther}}]{kinos21}%
  \BibitemOpen
  \bibfield  {author} {\bibinfo {author} {\bibfnamefont {A.}~\bibnamefont
  {Kinos}}, \bibinfo {author} {\bibfnamefont {D.}~\bibnamefont {Hunger}},
  \bibinfo {author} {\bibfnamefont {R.}~\bibnamefont {Kolesov}}, \bibinfo
  {author} {\bibfnamefont {K.}~\bibnamefont {M{\o}lmer}}, \bibinfo {author}
  {\bibfnamefont {H.}~\bibnamefont {{de Riedmatten}}}, \bibinfo {author}
  {\bibfnamefont {P.}~\bibnamefont {Goldner}}, \bibinfo {author} {\bibfnamefont
  {A.}~\bibnamefont {Tallaire}}, \bibinfo {author} {\bibfnamefont
  {L.}~\bibnamefont {Morvan}}, \bibinfo {author} {\bibfnamefont
  {P.}~\bibnamefont {Berger}}, \bibinfo {author} {\bibfnamefont
  {S.}~\bibnamefont {Welinski}}, \bibinfo {author} {\bibfnamefont
  {K.}~\bibnamefont {Karrai}}, \bibinfo {author} {\bibfnamefont
  {L.}~\bibnamefont {Rippe}}, \bibinfo {author} {\bibfnamefont
  {S.}~\bibnamefont {Kr{\"o}ll}},\ and\ \bibinfo {author} {\bibfnamefont
  {A.}~\bibnamefont {Walther}},\ }\href@noop {} {\bibfield  {journal} {\bibinfo
   {journal} {arXiv:2103.15743 [quant-ph]}\ } (\bibinfo {year} {2021})},\
  \Eprint {https://arxiv.org/abs/2103.15743} {arXiv:2103.15743 [quant-ph]}
  \BibitemShut {NoStop}%
\end{thebibliography}
%

\end{document}